\documentclass[12pt,a4paper]{article}

\usepackage{subfigure}
\usepackage[nosort]{cite}
\usepackage[hyperref,bulletsep]{collect}
\usepackage{bbm}
\usepackage{amsmath}
\usepackage{amssymb}
\usepackage{amscd}
\usepackage{slashed}
\usepackage{array}
\usepackage{rotating}
\usepackage{float}
\usepackage{subfigure}
\usepackage{feynmp}

\makeatletter \@addtoreset{equation}{section} \makeatother

\makeatletter
\let\old@startsection=\@startsection
\let\oldl@section=\l@section
\renewcommand{\@startsection}[6]{\old@startsection{#1}{#2}{#3}{#4}{#5}{#6\mathversion{bold}}}
\renewcommand{\l@section}[2]{\oldl@section{\mathversion{bold}#1}{#2}}
\makeatother

\makeatletter
\let\old@makecaption=\@makecaption
\def\@makecaption{\small\old@makecaption}
\makeatother

\renewcommand{\thefootnote}{\arabic{footnote}}
\let\oldPhi=\Phi
\let\oldPsi=\Psi
\let\oldGamma=\Gamma
\let\oldDelta=\Delta
\let\oldSigma=\Sigma
\let\oldTheta=\Theta
\let\oldPi=\Pi
\let\oldUpsilon=\Upsilon
\renewcommand{\Phi}{\mathnormal{\oldPhi}}
\renewcommand{\Psi}{\mathnormal{\oldPsi}}
\renewcommand{\Gamma}{\mathnormal{\oldGamma}}
\renewcommand{\Sigma}{\mathnormal{\oldSigma}}
\renewcommand{\Delta}{\mathnormal{\oldDelta}}
\renewcommand{\Theta}{\mathnormal{\oldTheta}}
\renewcommand{\Pi}{\mathnormal{\oldPi}}
\renewcommand{\Upsilon}{\mathnormal{\oldUpsilon}}

\newcommand{\tr}{\mathop{\mathrm{tr}}}

\ifx\genfrac\sdflkaj

\else

\fi

\newcommand{\bigbrk}[1]{\bigl(#1\bigr)}

\newcommand{\Bigsbrk}[1]{\Bigl[#1\Bigr]}

\newcommand{\vev}[1]{\langle#1\rangle}

\newcommand{\nl}[1][0pt]{\nonumber\\[#1]&\hspace{-4\arraycolsep}&\mathord{}}

\newcommand{\earel}[1]{\mathrel{}&\hspace{-2\arraycolsep}#1\hspace{-2\arraycolsep}&\mathrel{}}
\newcommand{\eq}{\earel{=}}

\def\[{\begin{equation}}
\def\]{\end{equation}}

\newcommand{\be}{\begin{eqnarray}}
\newcommand{\ee}{\end{eqnarray}}

\makeatletter
\def\mr@ignsp#1 {\ifx\:#1\@empty\else #1\expandafter\mr@ignsp\fi}%
\newcommand{\multiref}[1]{\begingroup
\xdef\mr@no@sparg{\expandafter\mr@ignsp#1 \: }%
\def\mr@comma{}%
\@for\mr@refs:=\mr@no@sparg\do{\mr@comma\def\mr@comma{,}\ref{\mr@refs}}%
\endgroup}
\makeatother

\newcommand{\hypref}[2]{\ifx\href\asklfhas #2\else\href{#1}{#2}\fi}

\renewcommand{\eqref}[1]{(\multiref{#1})}

\ifx\href\asklfhas\newcommand{\href}[2]{#2}\fi

\newcommand{\levi}{\epsilon}

\newcommand{\deriD}{\mathcal{D}}

\begin{document}

\thispagestyle{empty}
\begin{flushright}\footnotesize
\texttt{arXiv:1002.0841}\\
\texttt{UUITP-04/10}\vspace{10mm}
\end{flushright}

\renewcommand{\thefootnote}{\fnsymbol{footnote}}
\setcounter{footnote}{0}

\begin{center}
{\Large\textbf{\mathversion{bold}
On thermodynamics of $\mathcal{N}=6$ superconformal Chern-Simons theory
}\par}

\vspace{1.5cm}

25 May 2010

\vspace{1.0cm}

\textrm{\bf Mikael Smedb{\"a}ck} 
\textrm{
\\
\vspace{2mm}
Department of Physics and Astronomy\\
Uppsala University, SE-751 20 Uppsala, Sweden }
\vspace{2mm}
\texttt{\\ mikael.smedback@fysast.uu.se}

\par\vspace{14mm}

\textbf{Abstract} \vspace{5mm}

\begin{minipage}{14cm}

We study thermodynamics of $\mathcal{N}=6$ superconformal Chern-Simons theory by computing quantum corrections to the free energy.
We find that in weakly coupled ABJM theory on 
$\mathbb{R}^2 \times S^1$, the leading correction is non-analytic in the 't Hooft coupling $\lambda$, and is approximately of order 
$\lambda^2 \log(\lambda)^3$.
The free energy is expressed in terms of the scalar thermal mass $m$, which is generated by screening effects.
We show that this mass vanishes to 1-loop order. We then go on to 2-loop order where we find a finite and positive mass squared $m^2$. 
We discuss differences in the calculation between Coulomb and Lorentz gauge.
Our results indicate that
the free energy is a monotonic function in $\lambda$ which interpolates smoothly to the
$N^{3/2}$
behaviour at strong coupling.

\end{minipage}

\end{center}

\vspace{1.5cm}

\newpage

\setcounter{page}{1}
\renewcommand{\thefootnote}{\arabic{footnote}}
\setcounter{footnote}{0}

\hrule
\tableofcontents
\vspace{8mm}
\hrule
\vspace{4mm}

\setlength{\extrarowheight}{5pt}





\section{Introduction}
Realizations of the $AdS/CFT$ duality 
\cite{Maldacena:1997re,Gubser:1998bc,Witten:1998qj}
exist in several different dimensions.
Recently,
great progress has been made on understanding the gauge theory side of the $AdS_4/CFT_3$ version of this duality.
The original upsurge in interest was
generated by the work by Bagger, Lambert and Gustavsson \cite{Bagger:2006sk,Bagger:2007jr,Bagger:2007vi,Gustavsson:2007vu}, now known as the BLG theory. The search for a description of the field theory side of the duality, i.e. a world-volume theory on M2-branes has a long history, and various obstacles had to be overcome.
For example, Schwarz had shown that it was impossible to preserve the right symmetries in Chern-Simons theories based on $U(N)$ gauge groups
\cite{Schwarz:2004yj}.
However, Bagger and Lambert managed to write down a Lagrangian with all the right symmetries: superconformal symmetry $OSp(8|4)$ and parity invariance. In the original formulation, it was a Chern-Simons theory based on an algebraic construct known as a ``3-algebra'', later reformulated as an ordinary quiver gauge theory by van Raamsdonk \cite{VanRaamsdonk:2008ft} (a possibility which was already discussed by Bandres et al. \cite{Bandres:2008vf}). However, this theory, impressive as it was, turned out not to describe more than (at most) two M2-branes\cite{Lambert:2008et,Distler:2008mk}.
In response, various attempts were made to generalize the BLG theory \cite{Gran:2008vi,Gomis:2008uv,Benvenuti:2008bt,Ho:2008ei,Bandres:2008kj,Gomis:2008be}.

This background set the stage for the breakthrough paper \cite{Aharony:2008ug}, which successfully generalized the set-up to an arbitrary number of M2-branes. This theory, now known as the ABJM model, is an $\mathcal{N}=6$ superconformal Chern-Simons field theory. It builds on previous work on superconformal Chern-Simons-matter models \cite{Gaiotto:2007qi,Gaiotto:2008sd,Hosomichi:2008jd}, but with enlarged supersymmetry.
Crucially, the superpotential in this formulation, while reducing to the BLG superpotential for gauge groups of rank 2, did not suffer from the same obstructions as the one used to define the BLG model. In particular, the ABJM model allows arbitrary ranks $N$ of the gauge group, circumventing the severe gauge group restrictions that unitarity placed on the BLG model
\cite{Gauntlett:2008uf,Papadopoulos:2008sk}
. The paper \cite{Aharony:2008ug} includes an analysis of the moduli space for
$U(N) \times U(N)$ gauge groups, showing that 
$N$ M2-branes sitting on a space with a $\mathbb{Z}_k$ singularity
is in fact a consistent dual interpretation of the theory.
Various checks of this interpretation have been carried out. The conjectured $\mathcal{N}=6$ supersymmetry of the model was promptly confirmed
\cite{Benna:2008zy,Bandres:2008ry}.
Progress has been made on understanding the role of monopole operators and their relation to the expected supersymmetry enhancement to $\mathcal{N}=8$ for Chern-Simons levels $k=1,2$ \cite{Berenstein:2008dc,Klebanov:2008vq,Park:2008bk,Imamura:2009ur,Gaiotto:2009tk,SheikhJabbari:2009kr,Benna:2009xd,Gustavsson:2009pm,Berenstein:2009sa,Kwon:2009ar,Kim:2009ia,Imamura:2009hc,Kwon:2010ev}.
Calculations of the superconformal index
match between the strong and weak coupling regions
\cite{Bhattacharya:2008bja,Kim:2009wb}.
Relations back to the original BLG theory have been established in some cases \cite{Lambert:2010ji}.
There is even mounting evidence that the ABJM model is integrable
in the planar limit
\cite{Nishioka:2008gz,Minahan:2008hf,Gaiotto:2008cg,Grignani:2008is,Gromov:2008bz,Gromov:2008qe,Ahn:2008aa,Bak:2008cp,McLoughlin:2008he,Kristjansen:2008ib,Minahan:2009te,Berenstein:2009qd,Bak:2009mq,Bak:2009tq,Minahan:2009aq,Minahan:2009wg}.
Chern-Simons theories often arise in condensed matter systems, and possible applications include recent studies of the integer and fractional quantum Hall effect and Hall transitions 
\cite{KeskiVakkuri:2008eb,Davis:2008nv,Fujita:2009kw,Hikida:2009tp,Alanen:2009cn}
and superconducting M2-branes
\cite{Gomis:2008vc,Gauntlett:2009zw,Denef:2009tp,Gauntlett:2009dn,Gauntlett:2009bh,Bak:2010yd}.

In this note we will study the ABJM model on $\mathbb{R}^2 \times S^1$ at weak coupling and finite temperature. Hence, we begin by introducing the ABJM model at finite temperatature in section
\ref{sec_abjm}.
In the dual picture, finite temperature corresponds to a black hole geometry
\cite{Witten:1998zw}.
Thermodynamic properties of the theory are most succinctly captured by the free energy.
This object has been extensively studied in the
$d=4$, $\mathcal{N}=4$ super Yang-Mills setting, both at weak 
\cite{Fotopoulos:1998es}
and strong \cite{Gubser:1996de,Klebanov:1996un}
coupling, including corrections in inverse powers of the coupling 
\cite{Gubser:1998nz}.
At weak coupling, the next to leading order correction
is due to
thermal screening
of gauge fields and scalars
\cite{VazquezMozo:1999ic,Kim:1999sg,Nieto:1999kc}.
One interesting question that we aim to study is 
whether the ABJM theory also exhibits thermal screening.

In the ABJM case, the strong coupling answer was calculated by Klebanov and Tseytlin \cite{Klebanov:1996un}, and a discussion of corrections 
can be found in \cite{Garousi:2008ik}.
On the gauge side, the free field theory result was obtained in \cite{Aharony:2008ug}. 
Comparing the results at strong and weak coupling, there are intriguing differences between the
$\mathcal{N}=4$ super Yang-Mills theory and the ABJM theory. 
In the ABJM case the gravity result is proportional to $N^{3/2}$, which is an artifact of M2-branes, while the field theory side has $N^2$ degrees of freedom.
Moreover, the entropy goes to a constant in the strong coupling limit in the $\mathcal{N}=4$ super Yang-Mills case, which is in stark contrast to the $\lambda^{-1/2}$ behaviour of the ABJM case.

One motivation for us is to 
make progress on understanding how these behaviours come about in ABJM theory.
Correspondingly, we compute the free energy, including the first non-vanishing quantum correction, in section
\ref{sec_perturbative}. In section \ref{sec_naive}, we show that a naive perturbation expansion does not work, which is closely related to the fact that the first non-vanishing quantum correction is indeed due to thermal screening of scalars, as we discuss in section \ref{sec_thermal_screening}. Instead, we need to use a method based on resummation of ring diagrams, as explained in section \ref{sec_reorganized}.
We then combine our results to write down the full free energy, including the leading correction, in section \ref{sec_coupling}. Our results are consistent with 
a free energy which interpolates smoothly to the $N^{3/2}$ behaviour at strong coupling.
We discuss this and other issues in section
\ref{sec_conclusions}, which also contains a short summary of our results.

This note contains three appendices. In appendix \ref{app_prop}, we list propagators and Feynman rules, and explain our notation. Appendix \ref{app_thermal} contains the details of the calculation of the thermal mass for the scalars. This calculation is a crucial part of this note, but has been moved to the appendix due to its lengthy and technical nature. The main gauge used throughout this note is Coulomb gauge, which is often very convenient for thermal calculations. Appendix \ref{app_gluons} illustrates some of the difficulties with Lorentz gauge, in particular the apparent non-existence of an IR regulating mass for the gluons, which are propagating in Lorentz gauge.


\section{ABJM at finite temperature}\label{sec_abjm}

The ABJM model is a three-dimensional Chern-Simons-matter theory with gauge groups
$U(N) \times U(N)$ and $SU(N) \times SU(N)$, with $\mathcal{N}=6$ superconformal symmetry\footnote{An introduction to the ABJM model is given in \cite{Klebanov:2009sg}.}.
Other choices of gauge groups exist, most notably those which are dual to orientifolded geometries \cite{Aharony:2008gk}.
At large N, the $U(N) \times U(N)$ theory is believed to be dual to M-theory on $AdS_4 \times S^7 / \mathbb{Z}_k$. 
Henceforth, we will restrict attention to the $U(N) \times U(N)$ theory. In this section, we will also define the ABJM model at finite temperature.

The ABJM model is defined by the following action\footnote{
For a discussion on pure spinor superfield 
formulations, see \cite{Cederwall:2008xu,Cederwall:2008vd}.}.

\be\label{eq_action}
  S \eq \frac{k}{2\pi} \int d^3x\: \Bigsbrk{ 
        \levi^{i j k} \tr \bigbrk{
        -\frac{i}{2} A_i \partial_j A_k + \tfrac{1}{3} A_i A_j A_k
        +\frac{i}{2} \hat{A}_i \partial_j \hat{A}_k - \tfrac{1}{3} \hat{A}_i \hat{A}_j \hat{A}_k
      }
\nl \hspace{20mm}
    + \tr (\deriD_i Y_A)^\dagger \deriD^i Y^A
    + i \tr \psi^{\dagger A} \slashed{\deriD} \psi_A
    + V^{\mathrm{bos}} + V^{\mathrm{ferm}}  
}.
\ee
We are using the notation of \cite{Benna:2008zy}, 
but we have rescaled all matter fields by
$Y^A \rightarrow \sqrt{\frac{k}{2\pi}} Y^A$ and 
$\psi_A \rightarrow \sqrt{\frac{k}{2\pi}}\psi_A$,
to be able to factor out the Chern-Simons level $k$ as an overall normalization of the action.
In addition, we have performed a Wick rotation to Euclidean space.
The Dirac matrices are
$\left( \gamma^i \right)_\alpha^{\hspace{2mm}\beta} =\left( -\sigma^2, \sigma^1, \sigma^3 \right)$,
where $\sigma^i$ are the Pauli spin matrices.

\begin{figure}[t]
 \centering 
    \begin{picture}(0,0)
{\Large
    \put(55, 50){\makebox(0,0)[l]{$U(N)$}}
    \put(189, 50){\makebox(0,0)[l]{$U(N)$}}
    \put(-20, 50){\makebox(0,0)[l]{$A_i$}}
    \put(285, 50){\makebox(0,0)[l]{$\hat{A}_i$}}
    \put(115, 107){\makebox(0,0)[l]{$Y^A,\psi_A$}}
    \put(115, -2){\makebox(0,0)[l]{$Y^{\dagger}_A,\psi^{\dagger A}$}}
}
    \end{picture}
  \includegraphics{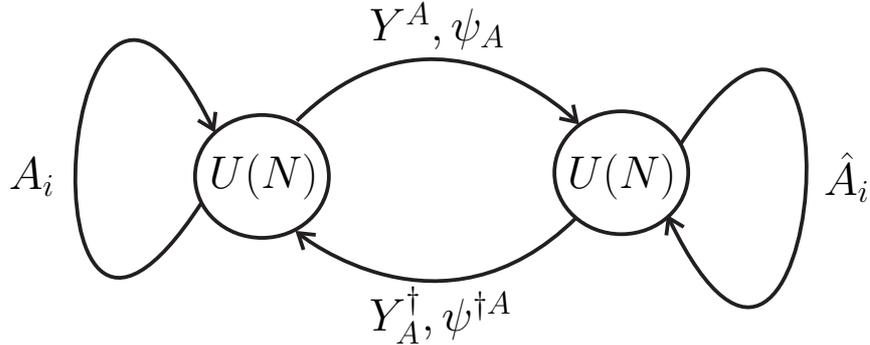}
  \caption{The ABJM model.}\label{fig_fieldcontent}
\end{figure}

The action (\ref{eq_action}) exhibits an $\mathcal{N}=6$ supersymmetry enhancement, owing to an underlying $SU(4)$ R-symmetry which acts on the matter fields $Y^A$ and $\psi_A$ ($A=1,2,3,4$).
As shown in \cite{Aharony:2008ug,Benna:2008zy},
this symmetry can be made manifest in the bosonic and fermionic potentials by writing them on the form
\be
   V^{\text{bos}} \eq - \frac{1}{3} \tr \Bigsbrk{
          Y^A Y_A^\dagger Y^B Y_B^\dagger Y^C Y_C^\dagger 
      +   Y_A^\dagger Y^A Y_B^\dagger Y^B Y_C^\dagger Y^C
\nl\hspace{11mm}
      + 4 Y^A Y_B^\dagger Y^C Y_A^\dagger Y^B Y_C^\dagger 
      - 6 Y^A Y_B^\dagger Y^B Y_A^\dagger Y^C Y_C^\dagger 
   } \; , \\
   V^{\text{ferm}} \eq i \tr \Bigsbrk{
        Y_A^\dagger Y^A \psi^{\dagger B} \psi_B
      - Y^A Y_A^\dagger \psi_B \psi^{\dagger B}
      + 2 Y^A Y_B^\dagger \psi_A \psi^{\dagger B}
      - 2 Y_A^\dagger Y^B \psi^{\dagger A} \psi_B
\nl\hspace{10mm}
      - \levi^{ABCD} Y_A^\dagger \psi_B Y_C^\dagger \psi_D
      + \levi_{ABCD} Y^A \psi^{\dagger B} Y^C \psi^{\dagger D}
      } \; .
\ee
The matter transforms in the bifundamental representation of the gauge group. Hence, the covariant derivative is
$
  \deriD_i Y^A = \partial_i Y^A + i A_i Y^A -i Y^A \hat{A}_i.
$
The gauge fields (henceforth referred to as gluons) $A_i$ and $\hat{A}_i$ ($i=1,2,3$) are in the adjoint of the left and right $U(N)$, respectively. The field content of the ABJM model is summarized in figure \ref{fig_fieldcontent}.

The action (\ref{eq_action}) has a $U(N) \times U(N)$ gauge-invariance, and needs to be gauge-fixed. 
We gauge-fix by adding
\be\label{eq_ghost}
  S_{\text{g}} \eq \frac{k}{2\pi} \int d^3x\: 
\left[
\tr \left( \frac{1}{2\xi}(\partial_\mu A^\mu)^2 + \frac{1}{2\hat{\xi}}(\partial_\mu \hat{A}^\mu)^2 \right)
- \tr \left( \bar{c} \; \partial_\mu \deriD^\mu \; c  +
 \bar{\hat{c}} \; \partial_\mu \deriD^\mu \; \hat{c} \right) \right]
\ee
to the action, $S \rightarrow S+S_{\text{g}}$. 
Two possible gauge choices are Lorentz gauge ($\mu=1,2,3$)
and Coulomb gauge ($\mu=1,2$). We typically use Coulomb gauge in the main text, and comment on the use of Lorentz gauge in footnotes.
$\xi$ and $\hat{\xi}$ are dimension one gauge-fixing parameters for $A$ and $\hat{A}$, and $c$ and $\hat{c}$ are the corresponding ghosts.
Hence, the covariant derivatives are $\deriD_\mu c = \partial_\mu c + i A_\mu c -i c A_\mu$,
and
$\deriD_\mu \hat{c} = \partial_\mu \hat{c} + i \hat{A}_\mu \hat{c} -i \hat{c} \hat{A}_\mu$.

To study the thermal theory, we compactify the time direction on a circle of length $L=1/T$, where $T$ is the temperature.
The action (\ref{eq_action}) and (\ref{eq_ghost}) then defines the theory on $\mathbb{R}^2 \times S^1$.
Finite temperature breaks both supersymmetry and conformal invariance.
A generic field $\phi$ has boundary condition
$\phi(x^1,x^2,x^3+L)=(-1)^{2\nu}\phi(x^1,x^2,x^3)$ along the compactified
direction.
In particular, the scalars, gluons and ghosts have periodic boundary conditions
($\nu=0$),
and the fermions are antiperiodic ($\nu=1/2$),
i.e.
\begin{equation}\label{eq_bc}
\begin{split}
  Y^A(x^1, x^2, x^3+L) & = + Y^A(x^1, x^2, x^3), \\
  \psi_A(x^1, x^2, x^3+L) & = - \psi_A(x^1, x^2, x^3), \\
  A_i(x^1, x^2, x^3+L) & = + A_i(x^1, x^2, x^3), \\
  c(x^1, x^2, x^3+L) & = +c(x^1, x^2, x^3).
\end{split}
\end{equation}
Propagators and vertices can
be derived from the action (\ref{eq_action}) and (\ref{eq_ghost}).
These Feynman rules are collected in appendix \ref{app_prop}. This appendix also contains an explanation of some of our notation.


\section{Perturbative free energy}\label{sec_perturbative}
We now want to compute the free energy, including the leading correction, and analyze its coupling dependence. We will show that to obtain a finite answer, we will need to reorganize perturbation theory, taking thermal screening of the scalars into account.


\subsection{Naive perturbation expansion}\label{sec_naive}

In this section, we will compute the free energy in a 
naive perturbation expansion.
We will show that perturbation theory breaks down
between two and three loops.

\begin{figure}[t]
 \centering 
\begin{equation} \nonumber
\raisebox{-8mm}{
\begin{fmffile}{diagF1loop}
    \begin{fmfgraph}(50,50)
       \fmfi{plain}{fullcircle scaled 1.2w shifted (-2.5w,.5h)}
       \fmfi{dashes}{fullcircle scaled 1.2w shifted (-.5w,.5h)}
       \fmfi{wiggly}{fullcircle scaled 1.2w shifted (1.5w,.5h)}
       \fmfi{dots}{fullcircle scaled 1.2w shifted (3.5w,.5h)}
    \end{fmfgraph}
\end{fmffile}}
\end{equation}
\put(-155, -2){\makebox(0,0)[l]{$(a)$}}
\put(-55, -2){\makebox(0,0)[l]{$(b)$}}
\put(45, -2){\makebox(0,0)[l]{$(c)$}}
\put(145, -2){\makebox(0,0)[l]{$(d)$}}
\caption{One-loop contributions to the free energy arise from (a) scalars and (b) fermions (and in Lorentz gauge, also from (c) gluons and (d) ghosts).}\label{fig_F1loop}
\end{figure}
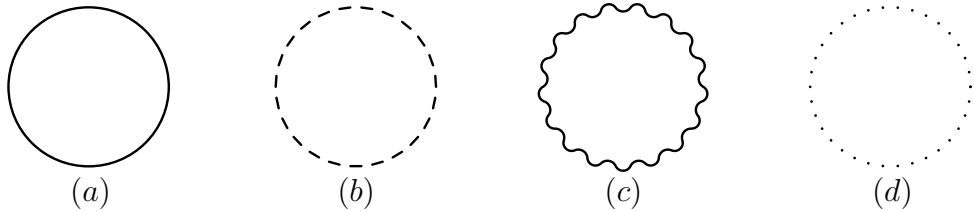

Assuming that the free energy is analytic
in the 't Hooft coupling $\lambda = \frac{N}{k}$,
we write
\begin{equation}\label{eq_perturbativeF}
  \tilde{F}(\lambda)=\tilde{F}_1 + \lambda \tilde{F}_2 + \lambda^2 \tilde{F}_3 + \cdots.
\end{equation}
We begin at one-loop order, using Coulomb gauge.
The free energy receives contributions from scalars and fermions
(see figure \ref{fig_F1loop}), which sum up to 
\begin{equation}\label{eq_Fabjm}
  \tilde{F}_1 = N^2 \left(8 \frac{1}{2}A_0-8\frac{1}{2}A_{1/2}\right) =
 -N^2 T^3 \frac{7 \zeta({3})}{\pi}.
\end{equation}
where
\begin{equation}\label{eq_A}
\begin{split}
  A_{\nu} & = \int \frac{d^2 p}{(2\pi)^2} T \sum_{n= \in \mathbb{Z}+\nu}  \log ( \vec{p}^2+\omega_n^2 ), \\
 A_{0}   & =-T^3 \frac{\zeta(3)}{\pi}, \\
  A_{1/2} & =T^3 \frac{3\zeta(3)}{4 \pi},
\end{split}
\end{equation}
corresponding to contributions from fields with
periodic ($\nu=0$) or antiperiodic ($\nu=1/2$)
boundary conditions.
Notice in (\ref{eq_Fabjm}) that the
matter fields each have eight real degrees of freedom.
Relative signs arise for the anticommuting fermions.
To verify that gluons and ghosts do not contribute, note that
\begin{equation}
  {\det}^{1/2}(\pm \levi^{ijk} \partial_k + \frac{1}{2\xi}\vec{\partial}_i \vec{\partial}_j)
  = \frac{\vec{\partial}^2}{\sqrt{2\xi}}.
\end{equation}
An arrow denotes that the object only has components along the $1,2$ directions, i.e. no dependence on $p_3$. Thus, this contribution vanishes in zeta function regularization. It is the Chern-Simons terms which make it possible to find a gauge in which gluons and ghosts are non-propagating\footnote{
In Lorentz gauge, there will also be contributions from gluons and ghosts, since
\begin{equation}
  {\det}^{1/2}(\pm \levi^{ijk} \partial_k + \frac{1}{2\xi}\partial_i \partial_j)
  = \frac{\partial^2}{\sqrt{2\xi}},
\end{equation}
so (\ref{eq_Fabjm}) becomes
\begin{equation}
  \tilde{F}_1 = N^2 \left(8 \frac{1}{2}A_0-8\frac{1}{2}A_{1/2}+2A_0-2A_0\right) =
 -N^2 T^3 \frac{7 \zeta({3})}{\pi}.
\end{equation}
Notice that the two types of gluons and ghosts (which are anticommuting, hence the relative sign) now do not vanish individually (as in Coulomb gauge), but instead cancel each other. In neither gauge is there a net contribution to the free energy from gluons and ghosts.
}.
In the zero-temperature limit $T \rightarrow 0$, the free energy goes to zero (independently of gauge choice).
At zero temperature (i.e. without compactifying the time direction), contributions from bosonic and fermionic degrees of freedom cancel automatically, as is required by supersymmetry.
The result (\ref{eq_Fabjm}) was already derived in \cite{Aharony:2008ug}.

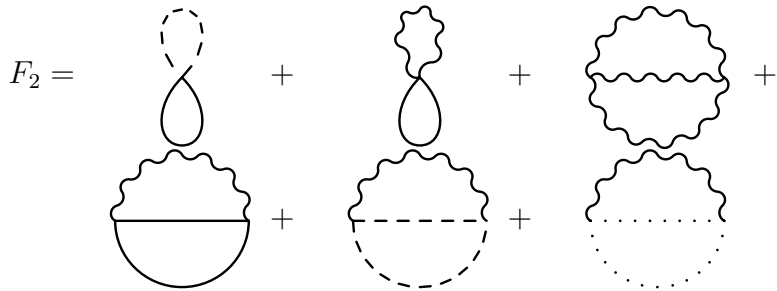
\begin{figure}[t]
  \centering 
    \begin{eqnarray}\nonumber
      F_2=\raisebox{-8mm}{
    \begin{fmffile}{diagF2loop1}
        \begin{fmfgraph}(50,50)
           \fmfleft{i}
           \fmfright{o}
           \fmf{phantom}{i,v}
           \fmf{phantom}{v,o}
           \fmf{dashes,tension=0.65}{v,v}
           \fmf{plain,left=90, tension=0.65}{v,v}
        \end{fmfgraph}
    \end{fmffile}}+
&
    \raisebox{-8mm}{
    \begin{fmffile}{diagF2loop2}
        \begin{fmfgraph}(50,50)
           \fmfleft{i}
           \fmfright{o}
           \fmf{phantom}{i,v,v,o}
           \fmf{wiggly,tension=0.65}{v,v}
           \fmf{plain,left=90, tension=0.65}{v,v}
        \end{fmfgraph}
    \end{fmffile}}+
&
 \raisebox{-8mm}{
    \begin{fmffile}{diagF2loop3}
        \begin{fmfgraph}(50,50)
           \fmfleft{i}
           \fmfright{o}
           \fmf{phantom}{i,v1}
           \fmf{phantom}{v2,o}
           \fmf{wiggly,left,tension=0}{v1,v2}
           \fmf{wiggly,tension=0}{v1,v2}
           \fmf{wiggly,right,tension=0}{v1,v2}
        \end{fmfgraph}
    \end{fmffile}}+
\\ \nonumber
 \raisebox{-8mm}{
    \begin{fmffile}{diagF2loop4}
        \begin{fmfgraph}(50,50)
           \fmfleft{i}
           \fmfright{o}
           \fmf{phantom}{i,v1}
           \fmf{phantom}{v2,o}
           \fmf{wiggly,left,tension=0}{v1,v2}
           \fmf{plain,tension=0}{v1,v2}
           \fmf{plain,right,tension=0}{v1,v2}
        \end{fmfgraph}
    \end{fmffile}}+
&
 \raisebox{-8mm}{
    \begin{fmffile}{diagF2loop5}
        \begin{fmfgraph}(50,50)
           \fmfleft{i}
           \fmfright{o}
           \fmf{phantom}{i,v1}
           \fmf{phantom}{v2,o}
           \fmf{wiggly,left,tension=0}{v1,v2}
           \fmf{dashes,tension=0}{v1,v2}
           \fmf{dashes,right,tension=0}{v1,v2}
        \end{fmfgraph}
    \end{fmffile}}+
&
\raisebox{-8mm}{
    \begin{fmffile}{diagF2loop6}
        \begin{fmfgraph}(50,50)
           \fmfleft{i}
           \fmfright{o}
           \fmf{phantom}{i,v1}
           \fmf{phantom}{v2,o}
           \fmf{wiggly,left,tension=0}{v1,v2}
           \fmf{dots,tension=0}{v1,v2}
           \fmf{dots,right,tension=0}{v1,v2}
        \end{fmfgraph}
    \end{fmffile}}
  \end{eqnarray}
\caption{Two-loop contributions to the free energy.}\label{fig_F2loop}
\end{figure}

Using the interactions (\ref{eq_interactions}) and propagators (\ref{eq_propagators}), we can now try to find the perturbative corrections to the free energy.
At two-loop order, the free energy receives contributions from the connected and one-particle-irreducible diagrams shown in figure \ref{fig_F2loop}. All the diagrams vanish individually, due to combinatorics. This is true both at zero temperature and finite temperature, and for both Coulomb and Lorentz gauge. Hence,
\begin{equation}\label{eq_F2}
  \tilde{F}_2=0.
\end{equation}

\begin{figure}[t]
  \centering 
 \raisebox{-8mm}{
    \begin{fmffile}{diagflower}
    \begin{fmfgraph}(150,150)
       \fmfleft{i1,i2,i3}
       \fmfright{o1,o2,o3}
       \fmf{phantom}{i1,v,o3}
       \fmf{phantom}{i2,v,o2}
       \fmf{phantom}{i3,v,o1}       
       \fmf{plain,right=90}{v,v}
       \fmf{plain,left=150}{v,v}
       \fmf{plain,left}{v,v}
       \fmfdot{v}
    \end{fmfgraph}
    \end{fmffile}}
\caption{A three-loop contribution to the free energy.}\label{fig_flower}
\end{figure}
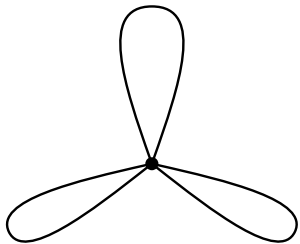

In view of (\ref{eq_F2}),
we might think that
the first non-vanishing correction terms 
to the free energy are of order 
$\mathcal{O}(\lambda^2)$.
However, this conclusion assumes that
the perturbation expansion (\ref{eq_perturbativeF}) is well-defined
around $\lambda=0$, and this is not true.
The obstruction consists of
infrared divergences, appearing
at three-loop order.
For example, the diagram shown in figure
\ref{fig_flower} is proportional to
\begin{equation}
  \left(  \int \frac{d^2 p}{(2\pi)^2}   T \sum_{n \in \mathbb{Z}} 
    \frac{1}{\vec{p}^2+\omega_n^2} \right)^3.
\end{equation}
The infrared divergences do not cancel. 
One possible way to regularize the theory is to go to finite volume. However, we are expressly interested in the non-compact theory, where the divergences should instead be cured by summing over ring diagrams. We will explain how this works in the following sections.


\subsection{Thermal screening}\label{sec_thermal_screening}

In section \ref{sec_naive},
we found that perturbation theory breaks down between two and three loops.
A similar phenomenon has been analyzed previously
in QCD
\cite{Arnold:1994ps,Arnold:1994eb,Kapusta:1979fh,Gross:1980br}
and in super Yang-Mills
\cite{VazquezMozo:1999ic,Kim:1999sg,Nieto:1999kc},
where scalars and gluons are screened by
quantum effects.
The ABJM theory exhibits a similar behaviour.

To proceed, we must reorganize perturbation theory
in a way which regularizes the infrared divergences
in the propagators
\cite{Arnold:1994ps,Arnold:1994eb}.
In momentum space, the quadratic
part of the action
(\ref{eq_action}) and (\ref{eq_ghost})
is
\begin{equation}\label{eq_S2}
\begin{split}
  S_{\text{2}} & = \frac{k}{2\pi} \int 
  \frac{d^2p}{(2\pi)^2} \; T \sum_{n \in \mathbb{Z}} \tr \text{{\Huge [}}
        \levi^{i j k}
          \left(
        \frac{1}{2} A_i(p) p_j A_k(-p)
        -\frac{1}{2} \hat{A}_i(p) p_j \hat{A}_k(-p)
          \right)
\\
& + 
  \frac{1}{2\xi} A_i(p) p^2 A_i(-p)
 +\frac{1}{2\hat{\xi}} \hat{A}_i(p) p^2 \hat{A}_i(-p) 
+
\bar{c}(p)  p_{\mu}p^{\mu} c(-p)  +
\bar{\hat{c}}(p) p_{\mu}p^{\mu}  \hat{c}(-p)
\\
  &  +  Y^\dagger_A(p) p^2 Y^A(-p)
\text{{\Huge ]}}
-
  \frac{k}{2\pi} \int 
  \frac{d^2p}{(2\pi)^2} \; T \sum_{n \in \mathbb{Z} + \frac{1}{2}} \tr \text{{\Huge [}}
    \psi^{\dagger A}(p) \; \slashed{p} \; \psi_A(-p)  
 \text{{\Huge ]}}.
\end{split}
\end{equation}
We now write the action as
$S = \left( S + \delta S_2 \right) - \delta S_2$,
where
\begin{equation}\label{eq_deltaS2}
\begin{split}
  \delta S_{\text{2}}  = \frac{k}{2\pi} \int 
  \frac{d^2p}{(2\pi)^2} \; T \sum_{n \in \mathbb{Z}}  \tr \text{{\Huge [}} 
& \frac{1}{2}
  Y^\dagger_A(p) m^2_{Y} Y^A(-p) \text{{\Huge ]}} .
\end{split}
\end{equation}
We henceforth treat $- \delta S_2$ 
as a perturbation to $\left( S + \delta S_2 \right)$.
Effectively, this means that
infrared divergences are regularized by the generation of a thermal mass $m_Y^2$ in the propagator for the scalars\footnote{
In other gauges, thermal masses may be generated for all fields with bosonic boundary conditions.
}:
\begin{equation}\label{eq_newYY}
 \vev{Y_n(\vec{p}) \; Y^\dagger_{-n}(-\vec{p})}
    = \frac{2\pi}{k T} \, \frac{1}{\vec{p}^2+\omega_n^2+m_Y^2} \; .
\end{equation}
The generation of thermal masses is similar to the renormalization of the electric charge
in QED, where virtual electron-positron pairs are created and
cause a dielectric screening effect from the vacuum.
A thermal mass $m$ 
corresponds to generation of a finite static screening length $r=\frac{1}{m}$
\cite{Gross:1980br}.

Technically, the scalar thermal mass is obtained by computing the
one-particle-irreducible contributions to the self-energy
in the static limit (the limit of no external momentum).
The details of the calculation (in Coulomb gauge) are collected in appendix \ref{app_thermal}.
The final result is given by equations 
(\ref{eq_scalar_selfenergy}) and
(\ref{eq_scalar_thermal_mass}),
quoted here for convenience:
\begin{equation}\label{eq_scalar_summary}
\begin{split}
  m_Y^2(\lambda) & = (2\pi T)^2\mu^2(\lambda), \\
  \mu^2(\lambda) 
& = \frac{118}{3 (2\pi)^2} 
  \lambda^2 \log(\mu)^2 
  + 
  \mathcal{O}(\lambda^2 \log(\lambda)).
\end{split}
\end{equation}
A few comments are in order.
Since the first non-vanishing corrections (i.e. at two loops) to the thermal mass
are themselves infrared divergent,
the thermal mass must be computed self-consistently
by requiring that successive higher order perturbative corrections
do not shift the pole in the propagator (cf. D'Hoker's original calculation for $\text{QCD}_3$ \cite{D'Hoker:1981us}). 
Thus, the diagrams will themselves depend on the regulator that we are trying to compute. Notice that the right-hand side of equation (\ref{eq_scalar_summary}) depends on $\mu^2(\lambda)$. However, to a good approximation (for very small $\lambda$), we can replace $\log(\mu)$ by $\log(\lambda)$. Notice that $m_Y^2$ is then approximately of order
$\lambda^2 \log(\lambda)^2$, rather than just $\lambda^2$.

In principle, it would have been useful to verify that we are indeed expanding around the correct vacuum of the finite temperature theory\footnote{
I thank E. Witten for drawing my attention to this issue.
}. Here, we will be content with the observation that the mass squared $m_Y^2$ is manifestly positive.
It would also have been useful to verify our results by an analogous treatment in Lorentz gauge. 
In Lorentz gauge, gluons are propagating, and would apparently need a regulator to yield a finite answer. However, no such IR regulating mass exists for the gluons up to two-loop order, as we show in appendix \ref{app_gluons}.


\subsection{Reorganized perturbation expansion}\label{sec_reorganized}
We have concluded that finding the leading correction to the free energy requires reorganizing perturbation theory.
Thus, we need to replace (\ref{eq_perturbativeF}) by
\begin{equation}\label{eq_Ftotal}
  F(\lambda)=F_1(\lambda) + \lambda F_2(\lambda) + \lambda^2 F_3(\lambda) \cdots,
\end{equation}
where $F_1(\lambda)$, $F_2(\lambda)$, $F_3(\lambda)$, $\ldots$
are now computed using the renormalized scalar propagator
(\ref{eq_newYY}).
There is a $\lambda$
dependence in $F_1(\lambda)$, $F_2(\lambda)$, 
$F_3(\lambda)$, $\ldots$.
The reason is that there is a $\lambda$ dependence in the thermal mass (\ref{eq_scalar_summary}).
In fact, it is this $\lambda$ dependence which gives rise to the first non-vanishing (and non-analytic) correction to the free energy, as we will now show.

\begin{figure}[t]
{\LARGE $\sum_{N}$} 
  \centering
 \raisebox{-25mm}{
    \begin{fmffile}{diagring}
    \begin{fmfgraph*}(150,150)
       \fmfleft{i1,i2}
       \fmfright{o1}
       \fmfv{decor.shape=circle,decor.filled=empty}{v1}
       \fmfv{decor.shape=circle,decor.filled=empty}{v2}
       \fmfv{decor.shape=circle,decor.filled=empty}{v3}
       \fmflabel{$1$ \hspace{0mm}}{v1}
       \fmflabel{$2$ \hspace{0mm}}{v2}
       \fmflabel{\hspace{0mm} $N$}{v3}
       \fmf{phantom}{i1,v1}
       \fmf{phantom}{i2,v2}
       \fmf{phantom}{o1,v3}
       \fmf{plain,tension=0.2,left=0.6}{v1,v2}
       \fmf{dots,tension=0.2,left=0.6}{v2,v3}
       \fmf{plain,tension=0.2,left=0.6}{v3,v1}
    \end{fmfgraph*}
    \end{fmffile}}
\caption{Free energy ring diagrams.}
\label{fig_ring}
\end{figure}
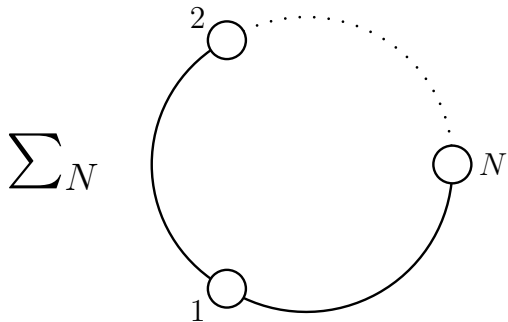


We begin with 
the resummed one-loop contribution $F_1(\lambda)$
(see figure \ref{fig_ring}).
Still working in Coulomb gauge, we find
\begin{equation}\label{eq_F1resummed}
  F_1(\lambda) = N^2 \left(8 \frac{1}{2}A_0(m_Y^2)-8\frac{1}{2}A_{1/2}(m_{\Psi}^2)\right),
\end{equation}
which generalizes the free field theory result (\ref{eq_Fabjm}) to the case of non-zero masses.
Fields with periodic boundary conditions ($\nu=0$) and mass $m$
contribute
\begin{align}\label{eq_Aperiodic}
  A_0(m^2) & = T \sum_{n\in\mathbb{Z}}  \int \frac{\text{d}^2 p}{(2 \pi)^2} \log{\left( \vec{p}^2+ \left(2\pi T n\right)^2+m^2\right)} = \\ \nonumber
& =\frac{T^3}{4\pi} \left[ -4 \zeta(3) - m^2 T^{-2} \log{(m^2T^{-2})}  
+m^2T^{-2}
+\mathcal{O}\left(m^4T^{-4}\right)
 \right],
\end{align}
and fields with antiperiodic boundary conditions ($\nu=1/2$)
contribute
\begin{align}\label{eq_Aantiperiodic}
  A_{1/2}(m^2)&=T \sum_{n\in\mathbb{Z}+\frac{1}{2}}  \int \frac{\text{d}^2 p}{(2 \pi)^2} \log{\left( \vec{p}^2 + \left( 2\pi T n \right)^2 + m^2\right)} = \\ \nonumber
& =\frac{T^3}{4\pi} \left[3 \zeta(3) 
-m^2T^{-2}\log(4)
+\mathcal{O}\left(m^4T^{-4}\right)
\right].
\end{align}
In the free field theory limit $\lambda \rightarrow 0$,
we recover the leading order results (\ref{eq_Fabjm}) and (\ref{eq_A}),
i.e. $\lim_{m\rightarrow0}A_{0}(m)=A_0$ and $\lim_{m\rightarrow0}A_{1/2}(m)=A_{1/2}$.
Note also that there is no zero mode for fields with antiperiodic boundary conditions, so no thermal mass will be generated for them. The dependence on $m^2$ in (\ref{eq_Aantiperiodic}) is analytic. Moreover, it follows from (\ref{eq_Aperiodic}) and (\ref{eq_Aantiperiodic}) that the leading order correction will anyway come from fields with periodic boundary conditions (i.e. the scalars).

The leading correction to 
the free energy is fully
contained in $F_1(\lambda)$.
Neither $F_2(\lambda)$ nor $F_3(\lambda)$
contributes, for the following reasons. $F_2(\lambda)$ still receives contributions from
the same diagrams as before, shown in
figure \ref{fig_F2loop}. Hence, it still vanishes for combinatorical reasons.
On the other hand, $F_3(\lambda)$ is not expected to vanish.
However, all contributions to the leading correction are already included in $F_1(\lambda)$, due to the definition of the thermal mass. 
Notice that we obtain 3-loop diagrams by closing the scalar propagators in the diagrams in figures
\ref{fig_644}, \ref{fig_433} and \ref{fig_433vanishing}.
Any such prospective additional contributions to $F_3(\lambda)$ will be cancelled by counterterms. Equivalently, D'Hoker's self-consistent treatment requires that the pole in the scalar propagator is not shifted by higher order corrections \cite{D'Hoker:1981us}. 


\subsection{Dependence on the coupling}\label{sec_coupling}
We have argued
that the leading correction to the free energy is non-analytic
in the coupling, due to thermal screening of scalars. 
We now combine these results to
analyze the coupling dependence of the free energy.

The free energy density, including the leading 
correction, is
\begin{equation}\label{eq_free_final}
  F=-N^2 T^3 f(\lambda),
\end{equation}
where
\begin{equation}\label{eq_f}
  f(\lambda)=\left[
    \frac{7 \zeta(3)}{\pi}
    + \frac{m_Y^2(\lambda)}{\pi T^{2}} 
    \log\left(\frac{m_Y^2(\lambda)}{ T^{2}} \right)
    + \mathcal{O}(\lambda^2 \log(\lambda)^2)
  \right],
\end{equation}
as follows from combining equations 
(\ref{eq_Ftotal}),
(\ref{eq_F1resummed}),
(\ref{eq_Aperiodic})
and
(\ref{eq_Aantiperiodic}),
and with the scalar thermal mass $m_Y^2(\lambda)$ defined by
equation (\ref{eq_scalar_summary}).
Using the approximation $\log(\mu) \approx \log(\lambda)$,
we get the convenient closed form answer
\begin{equation}\label{eq_fapprox}
  f_{\text{approx}}(\lambda)=\left[
    \frac{7 \zeta(3)}{\pi}
    + \frac{236}{3 \pi} \lambda^2 \log(\lambda)^3
  + \mathcal{O}(\lambda^2 \log(\lambda)^2)
  \right].
\end{equation}
We observe that the leading order
correction to the free energy is (approximately) of order
$\lambda^2 \log(\lambda)^3$.
Such a behaviour has also been observed 
in other three-dimensional Chern-Simons-matter theories with lower supersymmetry
\cite{Gaiotto:2007qi}.


\begin{figure}[t]
  \centering
  \subfigure[Free energy.]{\includegraphics[height=4cm]{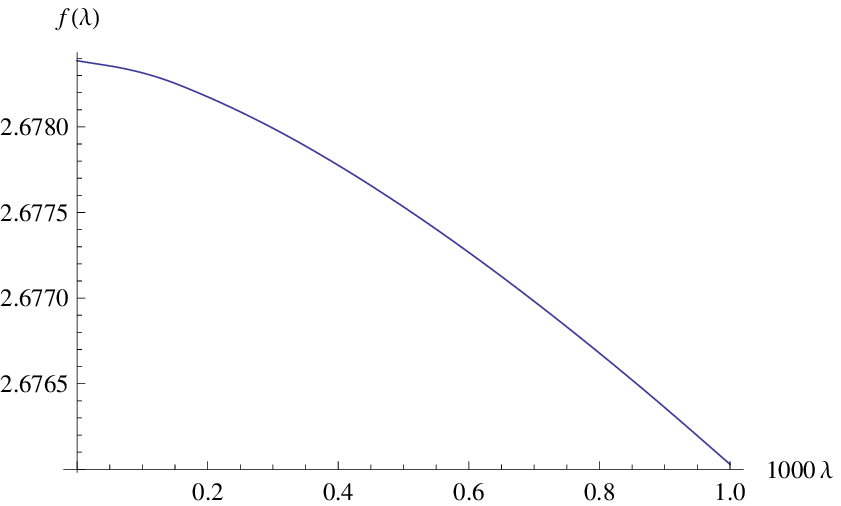}\label{fig_free}}
  \qquad
  \subfigure[Scalar thermal mass.]{\includegraphics[height=4cm]{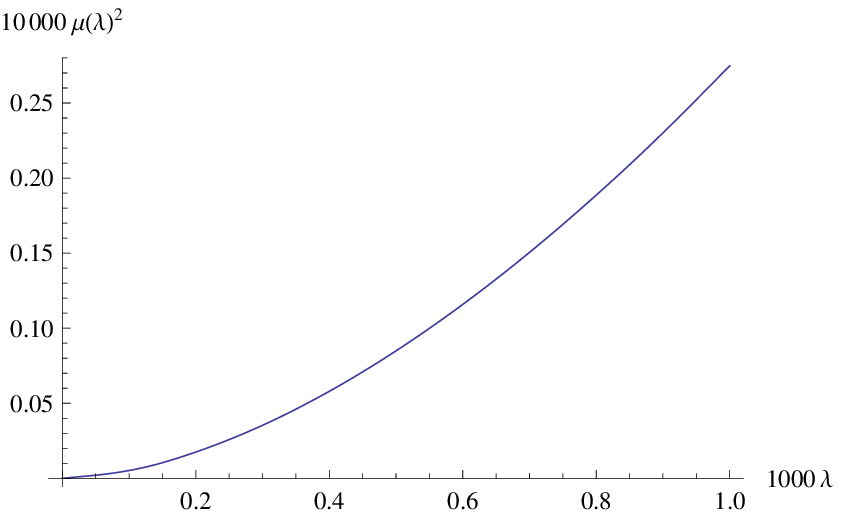}\label{fig_scalar}}
  \caption{Free energy and scalar thermal mass.}\label{fig_fs}
\end{figure}


Not using the approximation $\log(\mu) \approx \log(\lambda)$,
equation (\ref{eq_scalar_summary}) can be solved numerically.
The numerical solution, and the resulting free energy, are shown in table \ref{tab_fs} and plotted in figure \ref{fig_fs}. 
The solutions are not plotted 
beyond $\lambda = \exp(-7) \approx 0.001 $,
since the next to leading order correction will
be negligible only so long as $\log(\lambda)$ is sufficiently large.
The free field theory result for the free energy
is $f(0) \approx 2.6784$.

\begin{table}[htb]
  \centering
  \begin{tabular}{|c|c|c|}
    \hline
      $\lambda$ & $f(\lambda)$ & $10^4 \mu^2(\lambda)$ \\
    \hline
    \hline
0.0000 & 2.67839 & 0 \\
0.0001 & 2.67832 & 0.00521 \\
0.0002 & 2.67818 & 0.01751 \\
0.0003 & 2.67799 & 0.03533 \\ 
0.0004 & 2.67778 & 0.05795 \\
0.0005 & 2.67753 & 0.08490 \\
0.0006 & 2.67727 & 0.11584 \\
0.0007 & 2.67698 & 0.15049 \\
0.0008 & 2.67668 & 0.18864 \\
0.0009 & 2.67636 & 0.23011 \\
0.0010 & 2.67603 & 0.27473 \\
    \hline
  \end{tabular}
  \caption{Free energy and scalar thermal mass.}\label{tab_fs}
\end{table}

We see that our results indicate that
the free energy is a monotonic function in $\lambda$ which interpolates smoothly to the
$N^{3/2}$
behaviour at strong coupling, where the free energy is given by\footnote{
For comparison, in $\mathcal{N}=4$ SYM the analogous results in the weak and strong coupling regions are \cite{Fotopoulos:1998es,Gubser:1998nz}
\begin{equation}\nonumber
  F_{\text{SYM}}(\lambda) = -\frac{\pi^2}{6} N^2 V_3 T^4 g(\lambda),
\end{equation}
\begin{equation}\nonumber
\begin{split}
  g_{\text{\text{weak}}}(\lambda) & = \left[ 1-\frac{3}{2\pi^2} \lambda + \cdots \right],  \\
  g_{\text{\text{strong}}}(\lambda) & = \left[ \frac{3}{4} + \frac{45}{32} \zeta(3)\frac{1}{\lambda^{3/2}} + \cdots \right].
\end{split}
\end{equation}
}
\cite{Klebanov:1996un}
\begin{equation}
  f_{\text{s}}(\lambda) = \left[ \frac{2^{7/2}}{9} \pi^2 \frac{1}{\sqrt{\lambda}} + \cdots \right].
\end{equation}


\section{Discussion}\label{sec_conclusions}
In this note, we have computed the free energy in ABJM theory on $\mathbb{R}^2 \times S^1$, given in equations
(\ref{eq_free_final}) and (\ref{eq_f}). The free energy is expressed in terms of the scalar thermal mass (\ref{eq_scalar_summary}), whose generation is due to screening effects. The numerical solution is shown in table \ref{tab_fs} and plotted in figure \ref{fig_fs}. Our answer for the free energy includes the first non-vanishing quantum correction, which is approximately of order $\lambda^2 \log(\lambda)^3$ (see equation (\ref{eq_fapprox})). The reason for the non-analytical dependence on the coupling is that the IR divergences had to be cured by using a technique based on resummation of ring diagrams.

Interesting differences appear compared to the calculation in $\mathcal{N}=4$ super-Yang Mills theory (SYM) \cite{Fotopoulos:1998es,VazquezMozo:1999ic,Kim:1999sg,Nieto:1999kc}. One important difference is that in SYM, the first non-vanishing correction is analytic in the coupling, and only after that do the non-analytic contributions appear. Another difference is that the self-consistent treatment is not necessary, since the 1-loop diagrams which define the thermal mass are regular. For us, already the first non-vanishing correction is non-analytic in the coupling, and the self-consistent treatment is required (note that the right-hand side of 
(\ref{eq_scalar_summary}) explicitly depends on the mass itself).

It would be interesting to compute higher order corrections to the free energy. However, even in the reorganized theory, the perturbation 
expansion is eventually expected to break down (at finite temperature)
\cite{Kapusta:1989bd}.
A related observation is that there are some indications 
\cite{Gross:1980br,Linde:1980ts} that even 
the self-consistent treatment proposed by D'Hoker
\cite{D'Hoker:1981us}, which we follow,
does not always eliminate all infrared divergences,
due to differences between electric and magnetic masses.
Our calculation of the free energy shows explicitly that the reorganized theory furnishes a finite answer to order $\lambda^2 \log(\lambda)^3$, but there is no guarantee that this finiteness persists to higher orders.

Another interesting question is the issue of phase transitions.
In SYM, there were some initial signs that the system may go through a phase transition between strong and weak coupling \cite{Li:1998kd,Gao:1998ww,Burgess:1999vb}. However, the present understanding is that the free energy is a smooth and monotonic function for all values of $\lambda$ \cite{Aharony:1999ti}, and more recent evidence also exists \cite{Blaizot:2006tk}.
However, even if the phase diagram on flat space is trivial in SYM, there are examples of three-dimensional field theories with possibly different behaviour, e.g. $\text{QCD}_3$ \cite{D'Hoker:1981us}. Nevertheless, we would expect ABJM to behave similarly to SYM in this respect.
Our results for the free energy and the scalar thermal mass are only valid so long as $\log(\lambda)$ is sufficiently large, but they still allow us to conjecture that $f(\lambda)$ is indeed smooth and monotonically decreasing between the weak and strong coupling regions also in ABJM theory.

A non-trivial phase space behaviour is expected on compact spaces. In the gravitational black hole description, the existence of a deconfinement (Hagedorn) transition has been established 
\cite{Witten:1998zw}.
SYM theory on a compact space was analyzed at zero 't Hooft coupling by Sundborg
by counting gauge-invariant states on a 3-sphere using P{\'o}lya theory \cite{Sundborg:1999ue}.
These results were verified and extended by
Aharony et al. \cite{Aharony:2003sx}, and later complemented by one-loop corrections \cite{Spradlin:2004pp}.
Similarly, using P{\'o}lya counting,
Nishioka and Takayanagi found a deconfinement transition in ABJM theory compactified on a 2-sphere \cite{Nishioka:2008gz}, and further study of such questions in ABJM would also be interesting.


\section*{Acknowledgements}

I thank Lisa Freyhult, Joseph Minahan, Olof Ohlsson Sax, Diego Rodriguez-Gomez, Edward Witten, Amos Yarom and in particular Igor Klebanov, Thomas Klose and Juan Maldacena for several very useful discussions. I thank Igor Klebanov and Thomas Klose for initial collaboration.
Most of this work was carried out while I was affiliated with and present at Princeton University.
This research was supported by
a Marie Curie Outgoing International Fellowship, contract No. MOIF-CT-2006-040369, within the 6th European Community Framework Programme.


\appendix

\section{Propagators and vertices}\label{app_prop}
In this section, we 
list propagators and interaction vertices,
which can be derived from the action
(\ref{eq_action}) and (\ref{eq_ghost}).

\begin{figure}[t]
  \centering 
\begin{eqnarray}\nonumber
    \raisebox{-8mm}{
    \begin{fmffile}{diaginteraction4gluon}
    \begin{fmfgraph}(50,50)
       \fmfleft{i1,i2}
       \fmfright{o1,o2}
       \fmf{plain}{i1,v1}
       \fmf{plain}{i2,v1}
       \fmf{wiggly}{o1,v1}
       \fmf{wiggly}{o2,v1}
       \fmfdot{v1}
    \end{fmfgraph}
    \end{fmffile}}
=V_4^{\text{g}}
& \hspace{10mm}
    \raisebox{-8mm}{
    \begin{fmffile}{diaginteraction4fermion}
    \begin{fmfgraph}(50,50)
       \fmfleft{i1,i2}
       \fmfright{o1,o2}
       \fmf{plain}{i1,v1}
       \fmf{plain}{i2,v1}
       \fmf{dashes}{o1,v1}
       \fmf{dashes}{o2,v1}
       \fmfdot{v1}
    \end{fmfgraph}
    \end{fmffile}}
=V_4^{\text{f}}
& \hspace{10mm}
    \raisebox{-8mm}{
    \begin{fmffile}{diaginteraction3gluon}
    \begin{fmfgraph}(50,50)
       \fmfleft{i1,i2,i3}
       \fmfright{o1,o2,o3}
       \fmf{wiggly}{i1,v1}
       \fmf{phantom}{i2,v1}
       \fmf{wiggly}{i3,v1}
       \fmf{phantom}{o1,v1}
       \fmf{wiggly}{o2,v1}
       \fmf{phantom}{o3,v1}
       \fmfdot{v1}
    \end{fmfgraph}
    \end{fmffile}}
=V_3^{\text{gl}}
\\ \nonumber
    \raisebox{-8mm}{
    \begin{fmffile}{diaginteraction3skalar}
    \begin{fmfgraph}(50,50)
       \fmfleft{i1,i2,i3}
       \fmfright{o1,o2,o3}
       \fmf{plain}{i1,v1}
       \fmf{phantom}{i2,v1}
       \fmf{plain}{i3,v1}
       \fmf{phantom}{o1,v1}
       \fmf{wiggly}{o2,v1}
       \fmf{phantom}{o3,v1}
       \fmfdot{v1}
    \end{fmfgraph}
    \end{fmffile}}
=V_3^{\text{s}}
& \hspace{10mm}
    \raisebox{-8mm}{
    \begin{fmffile}{diaginteraction3fermion}
    \begin{fmfgraph}(50,50)
       \fmfleft{i1,i2,i3}
       \fmfright{o1,o2,o3}
       \fmf{dashes}{i1,v1}
       \fmf{phantom}{i2,v1}
       \fmf{dashes}{i3,v1}
       \fmf{phantom}{o1,v1}
       \fmf{wiggly}{o2,v1}
       \fmf{phantom}{o3,v1}
       \fmfdot{v1}
    \end{fmfgraph}
    \end{fmffile}}
=V_3^{\text{f}}
& \hspace{10mm}
    \raisebox{-8mm}{
    \begin{fmffile}{diaginteraction3spoke}
    \begin{fmfgraph}(50,50)
       \fmfleft{i1,i2,i3}
       \fmfright{o1,o2,o3}
       \fmf{dots}{i1,v1}
       \fmf{phantom}{i2,v1}
       \fmf{dots}{i3,v1}
       \fmf{phantom}{o1,v1}
       \fmf{wiggly}{o2,v1}
       \fmf{phantom}{o3,v1}
       \fmfdot{v1}
    \end{fmfgraph}
    \end{fmffile}}
=V_3^{\text{gh}}
\\ \nonumber
& \hspace{10mm}
 \raisebox{-8mm}{
    \begin{fmffile}{diaginteraction6}
    \begin{fmfgraph}(50,50)
       \fmfleft{i1,i2,i3}
       \fmfright{o1,o2,o3}
       \fmf{plain}{i1,v1}
       \fmf{plain}{i2,v1}
       \fmf{plain}{i3,v1}
       \fmf{plain}{o1,v1}
       \fmf{plain}{o2,v1}
       \fmf{plain}{o3,v1}
       \fmfdot{v1}
    \end{fmfgraph}
    \end{fmffile}}
=V_6
\end{eqnarray}
\caption{Interaction vertices.}\label{fig_interactions}
\end{figure}
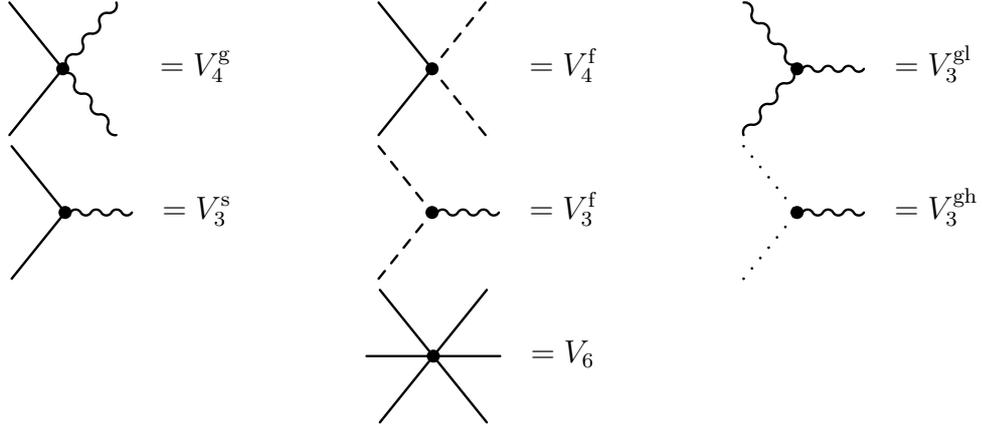

The interaction vertices are as follows\footnote{
In Lorentz gauge, $\mu$=1,2,3.
}
(see figure \ref{fig_interactions}).
\begin{equation}\label{eq_interactions}
\begin{split}
  V_6 &= - \frac{1}{3} \tr \Bigsbrk{
          Y^A Y_A^\dagger Y^B Y_B^\dagger Y^C Y_C^\dagger 
      +   Y_A^\dagger Y^A Y_B^\dagger Y^B Y_C^\dagger Y^C \\
      & \hspace{15mm} + 4 Y^A Y_B^\dagger Y^C Y_A^\dagger Y^B Y_C^\dagger 
      - 6 Y^A Y_B^\dagger Y^B Y_A^\dagger Y^C Y_C^\dagger 
   },
\\
  V_{4}^\text{g} &= \tr \Bigsbrk{ Y^A \hat{A}^i \hat{A}^i Y^{\dagger}_A
         + Y^{\dagger}_A A^i A^i Y^A
         - 2 \hat{A}^i Y^{\dagger}_A A^{i} Y^A },
\\
  V_{4}^\text{f} &= i \tr \Bigsbrk{
        Y_A^\dagger Y^A \psi^{\dagger B} \psi_B
      - Y^A Y_A^\dagger \psi_B \psi^{\dagger B}
      + 2 Y^A Y_B^\dagger \psi_A \psi^{\dagger B}
      - 2 Y_A^\dagger Y^B \psi^{\dagger A} \psi_B \\
      & \hspace{10mm} - \levi^{ABCD} Y_A^\dagger \psi_B Y_C^\dagger \psi_D
      + \levi_{ABCD} Y^A \psi^{\dagger B} Y^C \psi^{\dagger D}
      },
\\
  V_3^{\text{gl}}&= \frac{1}{3} \levi^{i j k} \tr \Bigsbrk{A_i A_j A_k - \hat{A}_i \hat{A}_j \hat{A}_k},
\\
  V_3^{\text{s}}&= i \tr \Bigsbrk{\partial_i Y^{\dagger}_A (A^i Y^A-Y^A \hat{A}^i) + \partial_i Y^A (\hat{A}^i Y^\dagger_A - Y^\dagger_A A^i)},
\\
  V_3^{\text{f}}&= \tr \Bigsbrk{-\psi^{\dagger A} A_i \gamma^i \psi_A + 
    \psi^{\dagger A} \gamma^i \psi_A \hat{A}_i},
\\
  V_3^{\text{gh}}&= i \tr \Bigsbrk{\partial_\mu \bar{c} (A^\mu c - c A^\mu) 
    + \partial_\mu \bar{\hat{c}} (\hat{A}^\mu \hat{c} - \hat{c} \hat{A}^\mu) }. \hspace{5mm} \text{(Coulomb gauge: $\mu=1,2$)}
\end{split}
\end{equation}
%
%
%
%
%
%
Using the Fourier transform
\begin{align}\label{app_fourier}
  & \int d^3 x \frac{e^{ipx}}{|x|^s} =c(s) \frac{1}{|p|^{3-s}} \; , \\ \nonumber
  & c(s) = 4 \pi \Gamma(2-s) \sin\left(\frac{\pi s}{2}\right) \; ,
\end{align}
we can derive the momentum space propagators (or read them off directly from (\ref{eq_S2}))
in Coulomb gauge\footnote{
In Lorentz gauge, the propagators for gluons and ghosts are
\begin{equation}
\begin{split}
  \vev{A_n(\vec{p}) \; A_{-n}(-\vec{p})} 
    & = -\frac{2\pi}{k T} \, \levi^{i j k} \frac{p_k}{\vec{p}^2+\omega_n^2}\; , \\
  \vev{\hat{A}_n(\vec{p}) \; \hat{A}_{-n}(-\vec{p})}
    & = +\frac{2\pi}{k T} \levi^{i j k} \frac{p_k}{\vec{p}^2+\omega_n^2} \; , \\
  \vev{c_n(\vec{p}) \; \bar{c}_{-n}(-\vec{p})}
     & = -\frac{2 \pi}{k T} \, \frac{1}{\vec{p}^2+\omega_n^2} \; , \\
  \vev{\hat{c}_n(\vec{p}) \; \bar{\hat{c}}_{-n}(-\vec{p})}
     & = -\frac{2\pi}{k T} \, \frac{1}{\vec{p}^2+\omega_n^2}.\;
\end{split}
\end{equation}
}:
\begin{equation}\label{eq_propagators}
\begin{split}
  \vev{Y_n(\vec{p}) \; Y^\dagger_{-n}(-\vec{p})}
    & = +\frac{2\pi}{k T} \, \frac{1}{\vec{p}^2+\omega_n^2} \; , \\
  \vev{\psi_{n}(\vec{p})_{\alpha} \; (\psi^\dagger_{-n})^{\beta}(-\vec{p})}
     & = -\frac{2\pi}{k T} \, \frac{\slashed{p}_{\alpha}^{\hspace{2mm} \beta}}{\vec{p}^2+\omega_n^2} \; , \\
  \vev{A_n(\vec{p}) \; A_{-n}(-\vec{p})} 
    & = -\frac{2\pi}{k T} \, (\levi^{i j 1} p_1 + \levi^{i j 2} p_2) \frac{1}{\vec{p}^2} \; , \\
  \vev{\hat{A}_n(\vec{p}) \; \hat{A}_{-n}(-\vec{p})}
    & = +\frac{2\pi}{k T} \, (\levi^{i j 1} p_1 + \levi^{i j 2} p_2) \frac{1}{\vec{p}^2} \; , \\
  \vev{c_n(\vec{p}) \; \bar{c}_{-n}(-\vec{p})}
     & = -\frac{2 \pi}{k T} \, \frac{1}{\vec{p}^2} \; , \\
  \vev{\hat{c}_n(\vec{p}) \; \bar{\hat{c}}_{-n}(-\vec{p})}
     & = -\frac{2\pi}{k T} \, \frac{1}{\vec{p}^2}. \;
\end{split}
\end{equation}
We use the Landau choice $\xi=\hat{\xi}=0$.
Delta functions in colour and flavour indices are suppressed.
In writing down the propagators, we use the
notation that $\vec{p}=(p_1,p_2)$, $(p_3)_n = \omega_n = 2 \pi T n$.
The modulus is $\sqrt{p^2}$,
where $p^2=\vec{p}^2+(\omega_n)^2$.
The index $n$ refers to different Fourier modes along the compactified direction. 
The mode expansion must adhere to the boundary conditions (\ref{eq_bc}).
Hence, 
$n \in \mathbb{Z} + \nu$, where
$\nu$ is $1/2$ for the fermions $\psi_A$, and otherwise zero.
For example, the scalars and the fermions are expanded as
\be
\begin{split}
  Y^A(x^1, x^2, x^3) & = T \sum_{n \in \mathbb{Z}} Y_n^A(x^1, x^2) e^{-i x^3 \omega_n}, \\
  \psi_A(x^1, x^2, x^3) & = T \sum_{n \in \mathbb{Z}+\frac{1}{2}} 
\left( \psi_{A}\right)_n(x^1, x^2) e^{-i x^3 \omega_n}.
\end{split}
\ee
We also make use of the symbols
\begin{equation}\label{eq_shortprop}
\begin{split}
  \Delta (p) & = \frac{1}{\vec{p}^2+\omega_n^2}, \\
  \Delta_S (p) & = \frac{1}{\vec{p}^2+\omega_n^2+m_Y^2}, \\
  \Delta_A (\vec{p}) & = \Delta_c (\vec{p}) = \frac{1}{\vec{p}^2}, 
\end{split}
\end{equation}
(where $p=(\vec{p},2 \pi T n)$),
corresponding to the generic, scalar, gluon and ghost propagator, respectively. 
Notice that $\Delta_S(p)$ includes the mass renormalization,
as in (\ref{eq_rens}).
The symbols $\Delta_A (\vec{p})$ and $\Delta_c(\vec{p})$ are equal, but it is still useful to remember their respective origins.
We use the subscript $f$ to denote fermionic boundary conditions. Explicitly, $p_f=(\vec{p},2 \pi T (n+1/2))$.
For example,
\begin{equation}\label{eq_not1}
\begin{split}
  \Delta (p_f) & = \frac{1}{\vec{p}^2+\omega_{n+1/2}^2}, \\
  \Delta (p_{f1}+p_{f2}) & = \frac{1}{(\vec{p_1}+\vec{p}_2)^2+\omega_{n_1+n_2+1}^2}.
\end{split}
\end{equation}
An arrow on top of an argument denotes
setting $p_3 \rightarrow 0$. For example,
\begin{equation}\label{eq_not2}
\begin{split}
  \Delta (\vec{p}) & = \frac{1}{\vec{p}^2}.
\end{split}
\end{equation}




\section{Scalar thermal mass}\label{app_thermal}

In this section, we compute the scalar thermal mass to leading order in Coulomb gauge.
Stated more precisely, our objective is to compute
the scalar self-energy in the limit of vanishing
external momentum
(the static limit), 
defined by the sum of all one-particle-irreducible diagrams,
\begin{equation}\label{eq_defM}
\begin{split}
\raisebox{-17mm}{
\begin{fmffile}{diagYYtotal}
    \begin{fmfgraph}(100,100)
       \fmfleft{i1}
       \fmfright{o1}
       \fmfv{decor.shape=circle,decor.filled=empty}{v1}
       \fmf{plain}{o1,v1,i1}
    \end{fmfgraph}
\end{fmffile}} 
=-T\lim_{\vec{p}\rightarrow 0} M^2(\vec{p},0)
=-2 \pi k T^3 \mu^2(\lambda).
\end{split}
\end{equation}
We will find that the one-loop answer vanishes. We therefore need to proceed to two loops to find a finite answer.

We begin with the one-loop calculation.
We will organize the computation in the following way.
There are two distinct types of diagrams which contribute at one-loop order.
These diagrams contain interaction terms of the types
\begin{align}\nonumber
  & \text{type $(4)$:} \hspace{10mm} \left( V_4^{\text{g}} + V_4^{\text{f}} \right), \\ \nonumber
  & \text{type $(33)$:} \hspace{9mm} (V_3^{\text{s}} + V_3^{\text{f}} + V_3^{\text{gl}} + V_3^{\text{gh}})^2,
\end{align}
in the notation of (\ref{eq_interactions}).
There are 9 distinct terms of type $(4)$
and 78 distinct terms of type $(33)$.
From (\ref{eq_interactions}) and (\ref{eq_propagators}),
it follows that in the static limit
(no external momentum), all one-particle irreducible diagrams vanish:

\begin{equation}\label{eq_YYcon1}
\raisebox{-17mm}{
\begin{fmffile}{diagYY1}
    \begin{fmfgraph}(100,100)
       \fmfleft{i1}
       \fmfright{o1}
       \fmf{plain}{o1,v1,i1}
       \fmf{dashes,right=0.5}{v1,v1}
       \fmfdot{v1}
    \end{fmfgraph}
\end{fmffile}}
\sim
q_i
\left( \gamma^i \right)_\alpha^{\hspace{2mm}\alpha} 
=0,
\end{equation}

\vspace{-20mm}

\begin{equation}\label{eq_YYcon2}
\raisebox{-17mm}{
\begin{fmffile}{diagYY2}
    \begin{fmfgraph}(100,100)
       \fmfleft{i1}
       \fmfright{o1}
       \fmf{plain}{o1,v1,i1}
       \fmf{wiggly,right=0.5}{v1,v1}
       \fmfdot{v1}
    \end{fmfgraph}
\end{fmffile}}
\sim
\levi^{ijk}\delta_{ij}
=0,
\end{equation}

\vspace{-20mm}

\begin{equation}\label{eq_YYcon3}
\raisebox{-17mm}{
\begin{fmffile}{diagYY3}
    \begin{fmfgraph}(100,100)
       \fmfleft{i1}
       \fmfright{o1}
       \fmf{plain}{o1,v1}
       \fmf{plain}{v2,i1}
       \fmf{wiggly,right=1.0,tension=0.75}{v1,v2}
       \fmf{plain,right=0.75,tension=0.75}{v2,v1}
    \end{fmfgraph}
\end{fmffile}}
\sim 
\levi^{ijk}q_i q_j = 0.
\end{equation}



Hence, we need to proceed to two-loop order.
At this order, four different types of diagrams contribute,
due to interaction terms
\begin{align}
  & \text{type $(6)$:} \hspace{11mm} V_6, \\
  & \text{type $(44)$:} \hspace{8mm}  
      \left(  V_4^{\text{g}} + V_4^{\text{f}}   \right)^2, \\
  & \text{type $(433)$:} \hspace{6mm}  \left(V_4^{\text{g}} + V_4^{\text{f}} \right)
    \left( V_3^{\text{s}} + V_3^{\text{f}} + V_3^{\text{gl}} + V_3^{\text{gh}} 
    \right)^2, \\
  & \text{type $(3333)$:} \hspace{4mm} \left( V_3^{\text{s}} + V_3^{\text{f}} 
      + V_3^{\text{gl}} + V_3^{\text{gh}} \right)^4,
\end{align}
still written in the notation of (\ref{eq_interactions}).
There are 4, 45, 702 and 1365 distinct terms of type
$(6)$, $(44)$, $(433)$ and $(3333)$, respectively.
%
%

\begin{figure}[H]
\centering 
  \begin{eqnarray}\nonumber
    \begin{fmffile}{diagscalar1}
      \raisebox{-8mm}{
        \begin{fmfgraph}(100,100)
           \fmfleft{i}
           \fmfright{o}
           \fmf{plain}{i,v1}
           \fmf{plain}{o,v1}
           \fmf{plain}{v1,v1}
           \fmf{plain,left=90}{v1,v1}
        \end{fmfgraph}}
      \raisebox{0mm}{
        \begin{fmfgraph}(100,100)
           \fmfleft{i1,i2,i3,i4}
           \fmfright{o1,o2,o3,o4}
           \fmf{phantom}{i2,v2,o2}
           \fmf{plain}{i1,v1,o1}
           \fmffreeze
           \fmf{dashes,tension=0.1,left}{v1,v2}
           \fmf{dashes,tension=0.1,right}{v1,v2}
           \fmf{plain,tension=0.7,right}{v2,v2}
        \end{fmfgraph}}
      \raisebox{0mm}{
        \begin{fmfgraph}(100,100)
           \fmfleft{i1,i2,i3,i4}
           \fmfright{o1,o2,o3,o4}
           \fmf{phantom}{i2,v2,o2}
           \fmf{plain}{i1,v1,o1}
           \fmffreeze
           \fmf{wiggly,tension=1,left}{v1,v2}
           \fmf{wiggly,tension=1,right}{v1,v2}
           \fmf{plain,tension=0.7,right}{v2,v2}
        \end{fmfgraph}}
    \end{fmffile}
\\ \nonumber
    \begin{fmffile}{diagscalar2}
      \raisebox{-8mm}{
        \begin{fmfgraph}(100,100)
           \fmfleft{i1}
           \fmfright{o1}
           \fmf{phantom}{i1,v1,v2,o1}
           \fmffreeze
           \fmf{plain}{i1,v1}
           \fmf{plain}{v2,o1}
           \fmf{wiggly,left,tension=0.2}{v1,v2}
           \fmf{plain}{v1,v2}
           \fmf{wiggly,right,tension=0.2}{v1,v2}
        \end{fmfgraph}}
      \raisebox{-8mm}{
        \begin{fmfgraph}(100,100)
           \fmfleft{i1}
           \fmfright{o1}
           \fmf{phantom}{i1,v1,v2,o1}
           \fmffreeze
           \fmf{plain}{i1,v1}
           \fmf{plain}{v2,o1}
           \fmf{plain,left,tension=0.2}{v1,v2}
           \fmf{dashes}{v1,v2}
           \fmf{dashes,right,tension=0.2}{v1,v2}
        \end{fmfgraph}}
       \raisebox{-8mm}{
        \begin{fmfgraph}(100,100)
           \fmfleft{i1}
           \fmfright{o1}
           \fmf{phantom}{i1,v1,v2,o1}
           \fmffreeze
           \fmf{plain}{i1,v1}
           \fmf{plain}{v2,o1}
           \fmf{plain,left,tension=0.2}{v1,v2}
           \fmf{wiggly}{v1,v2}
           \fmf{wiggly,right,tension=0.2}{v1,v2}
        \end{fmfgraph}}
    \end{fmffile}
  \end{eqnarray}
\caption{These type $(6)$ and $(44)$ diagrams are non-vanishing in both Coulomb and Lorentz gauge.}
\label{fig_644}
\end{figure}
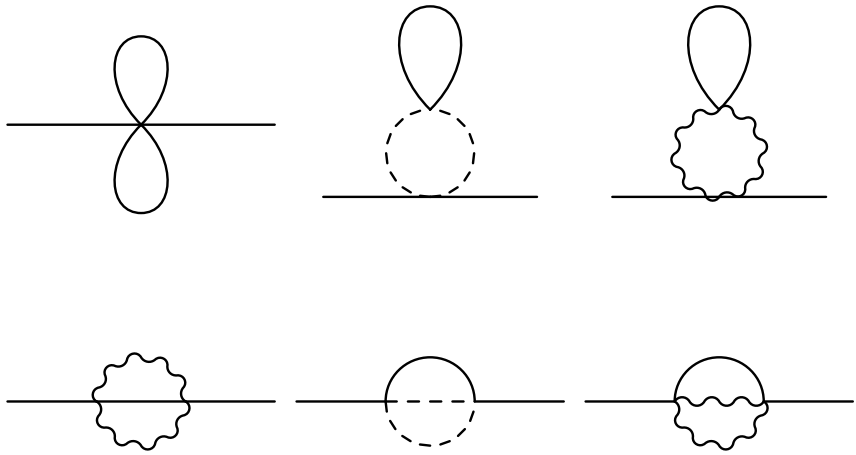
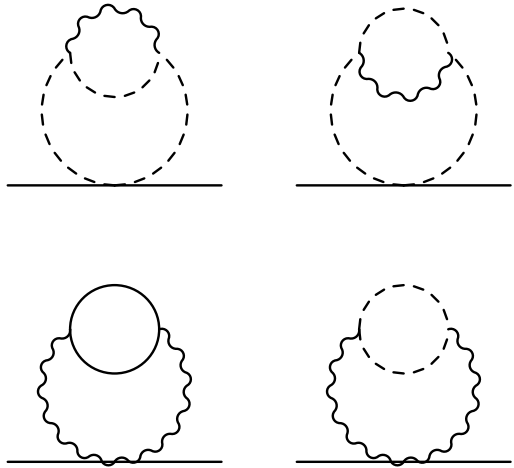
\begin{figure}[H]
\centering 
  \begin{eqnarray}\nonumber
    \begin{fmffile}{diagscalar3}
      \raisebox{-8mm}{
        \begin{fmfgraph}(100,100)
           \fmfleft{i1,i2,i3}
           \fmfright{o1,o2,o3}
           \fmf{phantom}{i2,v2,v3,o2}
           \fmf{plain}{i1,v1,o1}
           \fmffreeze
           \fmf{dashes,right=0.7,tension=2}{v1,v3}
           \fmf{dashes,left=0.7,tension=2}{v1,v2}
           \fmf{dashes,right}{v2,v3}
           \fmf{wiggly,left}{v2,v3}
        \end{fmfgraph}}
      \raisebox{-8mm}{
        \begin{fmfgraph}(100,100)
           \fmfleft{i1,i2,i3}
           \fmfright{o1,o2,o3}
           \fmf{phantom}{i2,v2,v3,o2}
           \fmf{plain}{i1,v1,o1}
           \fmffreeze
           \fmf{dashes,right=0.7,tension=2}{v1,v3}
           \fmf{dashes,left=0.7,tension=2}{v1,v2}
           \fmf{wiggly,right}{v2,v3}
           \fmf{dashes,left}{v2,v3}
        \end{fmfgraph}}
      \end{fmffile}
\\ \nonumber
    \begin{fmffile}{diagscalar4}
      \raisebox{-8mm}{
        \begin{fmfgraph}(100,100)
           \fmfleft{i1,i2,i3}
           \fmfright{o1,o2,o3}
           \fmf{phantom}{i2,v2,v3,o2}
           \fmf{plain}{i1,v1,o1}
           \fmffreeze
           \fmf{wiggly,right=0.7,tension=2}{v1,v3}
           \fmf{wiggly,left=0.7,tension=2}{v1,v2}
           \fmf{plain,right}{v2,v3}
           \fmf{plain,left}{v2,v3}
        \end{fmfgraph}}
      \raisebox{-8mm}{
        \begin{fmfgraph}(100,100)
           \fmfleft{i1,i2,i3}
           \fmfright{o1,o2,o3}
           \fmf{phantom}{i2,v2,v3,o2}
           \fmf{plain}{i1,v1,o1}
           \fmffreeze
           \fmf{wiggly,right=0.7,tension=2}{v1,v3}
           \fmf{wiggly,left=0.7,tension=2}{v1,v2}
           \fmf{dashes,right}{v2,v3}
           \fmf{dashes,left}{v2,v3}
        \end{fmfgraph}}
    \end{fmffile}
  \end{eqnarray}
\caption{These type $(433)$ diagrams are non-vanishing in both Coulomb and Lorentz gauge.}
\label{fig_433}
\end{figure}
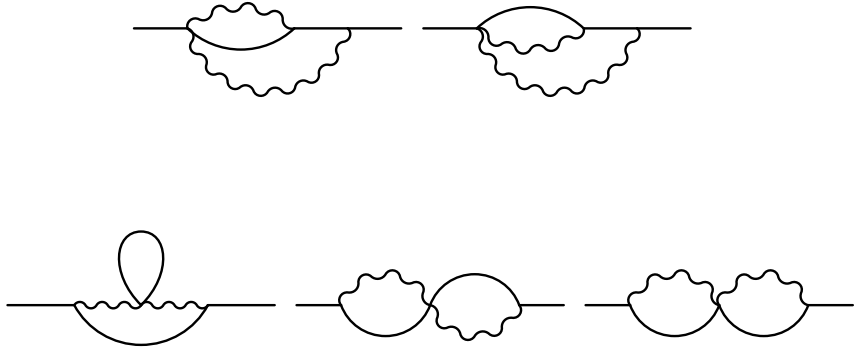
\begin{figure}[t]
\centering 
  \begin{eqnarray}\nonumber
    \begin{fmffile}{diagscalar5}
      \raisebox{-8mm}{
        \begin{fmfgraph}(100,100)
           \fmfleft{i}
           \fmfright{o}
           \fmf{phantom}{i,v1,v4,v2,v3,o}
           \fmffreeze
           \fmf{plain}{i,v1}
           \fmf{wiggly,left=0.4}{v1,v2}
           \fmf{plain,right=0.4}{v1,v2}
           \fmf{plain}{v2,v3}
           \fmf{wiggly,right=0.8}{v1,v3}
           \fmf{plain}{v3,o}
        \end{fmfgraph}}
      \raisebox{-8mm}{
        \begin{fmfgraph}(100,100)
\hspace{20mm}
           \fmfleft{i}
           \fmfright{o}
           \fmf{phantom}{i,v1,v4,v2,v3,o}
           \fmffreeze
           \fmf{plain}{i,v1}
           \fmf{plain,left=0.4}{v1,v2}
           \fmf{wiggly,right=0.4}{v1,v2}
           \fmf{plain}{v2,v3}
           \fmf{wiggly,right=0.8}{v1,v3}
           \fmf{plain}{v3,o}
        \end{fmfgraph}}
      \end{fmffile}
\\ \nonumber
    \begin{fmffile}{diagscalar6}
      \raisebox{-8mm}{
        \begin{fmfgraph}(100,100)
           \fmfleft{i}
           \fmfright{o}
           \fmf{phantom}{i,v1,v2,v3,o}
           \fmffreeze
           \fmf{plain}{i,v1}
           \fmf{wiggly}{v1,v2,v3}
           \fmf{plain,right=0.6}{v1,v3}
           \fmf{plain,right,tension=0.6}{v2,v2}
           \fmf{plain}{v3,o}
        \end{fmfgraph}}
      \raisebox{-8mm}{
        \begin{fmfgraph}(100,100)
           \fmfleft{i}
           \fmfright{o}
           \fmf{phantom}{i,v1,v2,v3,v4,v5,o}
           \fmffreeze
           \fmf{plain}{i,v1}
           \fmf{plain}{v5,o}
           \fmf{wiggly,left=0.7}{v1,v3}
           \fmf{plain,right=0.7}{v1,v3}
           \fmf{plain,left=0.7}{v3,v5}
           \fmf{wiggly,right=0.7}{v3,v5}
        \end{fmfgraph}}
      \raisebox{-8mm}{
        \begin{fmfgraph}(100,100)
           \fmfleft{i}
           \fmfright{o}
           \fmf{phantom}{i,v1,v2,v3,v4,v5,o}
           \fmffreeze
           \fmf{plain}{i,v1}
           \fmf{plain}{v5,o}
           \fmf{wiggly,left=0.7}{v1,v3}
           \fmf{plain,right=0.7}{v1,v3}
           \fmf{wiggly,left=0.7}{v3,v5}
           \fmf{plain,right=0.7}{v3,v5}
        \end{fmfgraph}}
    \end{fmffile}
  \end{eqnarray}
\caption{These type $(433)$ diagrams
vanish in Lorentz gauge (in the static limit) but not in Coulomb gauge.
}
\label{fig_433vanishing}
\end{figure}

We first consider a subset of the $(433)$ diagrams.
Thermal ABJM theory is severely plagued by IR divergences. However, there are some cancellations. In particular,
\begin{equation} \nonumber
    \begin{fmffile}{diagscalar7}
     \raisebox{-8mm}{
        \begin{fmfgraph}(100,100)
           \fmfleft{i1,i2,i3}
           \fmfright{o1,o2,o3}
           \fmf{phantom}{i2,v2,v3,o2}
           \fmf{plain}{i1,v1,o1}
           \fmffreeze
           \fmf{wiggly,right=0.7,tension=2}{v1,v3}
           \fmf{wiggly,left=0.7,tension=2}{v1,v2}
           \fmf{wiggly,right}{v2,v3}
           \fmf{wiggly,left}{v2,v3}
        \end{fmfgraph}} 
+
      \raisebox{-8mm}{
        \begin{fmfgraph}(100,100)
           \fmfleft{i1,i2,i3}
           \fmfright{o1,o2,o3}
           \fmf{phantom}{i2,v2,v3,o2}
           \fmf{plain}{i1,v1,o1}
           \fmffreeze
           \fmf{wiggly,right=0.7,tension=2}{v1,v3}
           \fmf{wiggly,left=0.7,tension=2}{v1,v2}
           \fmf{dots,right}{v2,v3}
           \fmf{dots,left}{v2,v3}
        \end{fmfgraph}}
      \end{fmffile} 
=
\end{equation}
\begin{align}\label{eq_cancel433gg}
=
  2\pi kT \lambda^2 
    \int \frac{d^2 q_1}{(2\pi)^2} T \sum_{n_1 \in \mathbb{Z}}  
    \int \frac{d^2 q_2}{(2\pi)^2} T \sum_{n_2 \in \mathbb{Z}} 
      I_g(q_1,q_2),
\end{align}
\begin{align}
\nonumber
I_g(q_1,q_2) =
& -\frac{4}{3} (\vec{q}_1 \cdot \vec{q}_2)^2 
\Delta_A(\vec{q}_1) \Delta_A(\vec{q}_2) 
\Delta_A(\vec{q}_1+\vec{q}_2)^2 \\ \nonumber
& +\frac{4}{3} \vec{q}_1^2 \vec{q}_2^2 \Delta_A(\vec{q}_1) \Delta_A(\vec{q}_2) \Delta_A(\vec{q}_1+\vec{q}_2)^2 \\ \nonumber
& +\frac{4}{3} (\vec{q}_1 \cdot \vec{q}_2)^2 
\Delta_c(\vec{q}_1) \Delta_c(\vec{q}_2) 
\Delta_A(\vec{q}_1+\vec{q}_2)^2 \\ \label{eq_cancel433gg_int}
& -\frac{4}{3} \vec{q}_1^2 \vec{q}_2^2 \Delta_c(\vec{q}_1) \Delta_c(\vec{q}_2) \Delta_A(\vec{q}_1+\vec{q}_2)^2 = 0, \nonumber
\end{align}
%
%
%
in the notation of (\ref{eq_shortprop}) and (\ref{eq_not2}).
Hence, all ghost contributions cancel already in
the integrand by equal but opposite gluon contributions.
Another way to see this is to notice that (\ref{eq_cancel433gg}) contains a piece which is constructed from 1-loop corrections to the gluon propagator.
This piece cancels by itself\footnote{
The same result holds in Lorentz gauge.
}:
\begin{equation}
\nonumber
\raisebox{-17mm}{
\begin{fmffile}{diagAAgluon}
    \begin{fmfgraph}(100,100)
       \fmfleft{i1}
       \fmfright{o1}
       \fmf{wiggly}{o1,v1}
       \fmf{wiggly}{v2,i1}
       \fmf{wiggly,right=0.75,tension=0.6}{v1,v2}
       \fmf{wiggly,right=0.75,tension=0.6}{v2,v1}
    \end{fmfgraph}
\end{fmffile}}
+
\raisebox{-17mm}{
\begin{fmffile}{diagAAghost}
    \begin{fmfgraph}(100,100)
       \fmfleft{i1}
       \fmfright{o1}
       \fmf{wiggly}{o1,v1}
       \fmf{wiggly}{v2,i1}
       \fmf{dots,right=0.9,tension=0.75}{v1,v2}
       \fmf{dots,right=0.9,tension=0.75}{v2,v1}
    \end{fmfgraph}
\end{fmffile}}
=
\end{equation}
\begin{equation}\nonumber
=
 2 k T \lambda 
    \int \frac{d^2 q}{(2\pi)^2} T \sum_{n \in \mathbb{Z}}  
  \left [  (\vec{q})_i (\vec{q})_j \Delta_A(\vec{q})^2
  - (\vec{q})_i (\vec{q})_j \Delta_c(\vec{q})^2
  \right]=0.
\end{equation}
%
We now consider all remaining diagrams.
One-particle-irreducible and non-vanishing diagrams contributing in the static limit are shown
in figures \ref{fig_644}, \ref{fig_433} and
\ref{fig_433vanishing}.
Adding together all contributions, we find\footnote{
In the flat case and for general momenta, the scalar self-energy to two loops was calculated in
\cite{Minahan:2009aq,Minahan:2009wg}.
}
\begin{equation}
-\lim_{\vec{p}\rightarrow 0} M^2(\vec{p},0)
=
  2\pi k \lambda^2 
    \int \frac{d^2 q_1}{(2\pi)^2} T \sum_{n_1 \in \mathbb{Z}}  
    \int \frac{d^2 q_2}{(2\pi)^2} T \sum_{n_2 \in \mathbb{Z}} 
      I(q_1,q_2),
\end{equation}
where
\begin{equation}
\begin{split}
  I(q_1,q_2) & = I_6(q_1,q_2) + I_{44}(q_1,q_2) + I_{433}(q_1,q_2) + 
  I_{3333}(q_1,q_2), \\
\\
  I_6(q_1,q_2) & = -42 \Delta_S(q_1) \Delta_S(q_2), \\
\\
  I_{44}(q_1,q_2) &= -32 \vec{q}_2^2 \Delta_S(q_1) \Delta_A(\vec{q}_2)^2 \\
         &  +4 (\vec{q}_1 \cdot \vec{q}_2) 
  \Delta_S(q_1) \Delta_A(\vec{q}_2) \Delta_A(\vec{q}_1+\vec{q}_2) \\
         &  +4 \vec{q}_2^2 \Delta_S(q_1) \Delta_A(\vec{q}_2) \Delta_A(\vec{q}_1+\vec{q}_2) \\
         &  -112 (q_{f1} \cdot q_{f2}) \Delta(q_{f1})
\Delta(q_{f2}) \Delta_S(q_{f1}+q_{f2}),
\end{split}
\end{equation}


%


\begin{equation}\nonumber
\begin{split}
  I_{433}(q_1,q_2) =
&-\frac{32}{3}  (\vec{q}_1 \cdot \vec{q}_2)^2
             \Delta(q_{f1}) \Delta(q_{f2}) 
             \Delta_A(\vec{q}_1+\vec{q}_2)^2 \\
&-\frac{64}{3} \vec{q}_1^2(\vec{q}_1 \cdot \vec{q}_2)
             \Delta(q_{f1}) \Delta(q_{f2}) 
             \Delta_A(\vec{q}_1+\vec{q}_2)^2 \\
&-\frac{32}{3} \vec{q}_1^2 \vec{q}_2^2
             \Delta(q_{f1}) \Delta(q_{f2}) 
             \Delta_A(\vec{q}_1+\vec{q}_2)^2 \\
&-\frac{2}{3} (q_1 \cdot q_2)
              (\vec{q}_1 \cdot \vec{q}_2)
             \Delta_A(\vec{q}_1) \Delta_A(\vec{q}_2)
             \Delta_S(q_1) \Delta_S(q_2) \\
&+\frac{2}{3} (\vec{q}_1 \cdot \vec{q}_2)^2
             \Delta_A(\vec{q}_1) \Delta_A(\vec{q}_2)
             \Delta_S(q_1) \Delta_S(q_2) \\
&+\frac{16}{3} q_2^2 \vec{q}_2^2
             \Delta_A(\vec{q}_2)^2
             \Delta_S(q_1) \Delta_S(q_2) \\
&-\frac{16}{3} \vec{q}_2^2 \vec{q}_2^2 
             \Delta_A(\vec{q}_2)^2
             \Delta_S(q_1) \Delta_S(q_2) \\
&-\frac{4}{3} (q_1 \cdot q_2)
              (\vec{q}_1 \cdot \vec{q}_2)
             \Delta_A(\vec{q}_2)                        
             \Delta_A(\vec{q}_1+\vec{q}_2) 
             \Delta_S(q_1) \Delta_S(q_2) \\
&-\frac{4}{3} (\vec{q}_1 \cdot \vec{q}_2)^2
             \Delta_A(\vec{q}_2)
             \Delta_A(\vec{q}_1+\vec{q}_2)
             \Delta_S(q_1) \Delta_S(q_2) \\
&-\frac{4}{3} q_2^2 (\vec{q}_1 \cdot \vec{q}_2)
             \Delta_A(\vec{q}_2)
             \Delta_A(\vec{q}_1+\vec{q}_2)
             \Delta_S(q_1) \Delta_S(q_2) \\
&-\frac{4}{3} \vec{q}_2^2 (q_1 \cdot q_2)
             \Delta_A(\vec{q}_2)
             \Delta_A(\vec{q}_1+\vec{q}_2)
             \Delta_S(q_1) \Delta_S(q_2) \\
&-\frac{4}{3} q_2^2 \vec{q}_2^2
             \Delta_A(\vec{q}_2)
             \Delta_A(\vec{q}_1+\vec{q}_2)
             \Delta_S(q_1) \Delta_S(q_2) \\
&+\frac{4}{3} \vec{q}_2^2 \vec{q}_2^2
             \Delta_A(\vec{q}_2)
             \Delta_A(\vec{q}_1+\vec{q}_2)
             \Delta_S(q_1) \Delta_S(q_2) \\
&+\frac{32}{3} (q_1 \cdot q_2)
               (\vec{q}_1 \cdot \vec{q}_2)
             \Delta_A(\vec{q}_1+\vec{q}_2)^2
             \Delta_S(q_1) \Delta_S(q_2) \\
&+\frac{32}{3} q_1^2(\vec{q}_1 \cdot \vec{q}_2)
             \Delta_A(\vec{q}_1+\vec{q}_2)^2
             \Delta_S(q_1) \Delta_S(q_2) \\
&+\frac{32}{3} \vec{q}_1^2(q_1 \cdot q_2)
             \Delta_A(\vec{q}_1+\vec{q}_2)^2
             \Delta_S(q_1) \Delta_S(q_2) \\
&+\frac{16}{3} q_1^2 \vec{q}_1^2
             \Delta_A(\vec{q}_1+\vec{q}_2)^2
             \Delta_S(q_1) \Delta_S(q_2) \\
&-\frac{16}{3} \vec{q}_1^2 \vec{q}_1^2
             \Delta_A(\vec{q}_1+\vec{q}_2)^2
             \Delta_S(q_1) \Delta_S(q_2) \\
&+\frac{16}{3} q_1^2 \vec{q}_2^2
             \Delta_A(\vec{q}_1+\vec{q}_2)^2
             \Delta_S(q_1) \Delta_S(q_2) \\
&+\frac{16}{3} \vec{q}_1^2 \vec{q}_2^2
             \Delta_A(\vec{q}_1+\vec{q}_2)^2
             \Delta_S(q_1) \Delta_S(q_2),
\end{split}
\end{equation}
in the notation of (\ref{eq_shortprop}), (\ref{eq_not1}) and (\ref{eq_not2}).
These expressions have a tendency to become more and more involved as the number of vertices increases. We omit spelling out the full expression for $I_{3333}(q_1,q_2)$. As we will see below, it has no zero modes and will not be of any further interest to us.

We expect the leading order result to
arise from the most IR divergent piece.
This piece (specifically, the piece proportional to 
$\lambda^2 \log(\mu)^2$) can be extracted from the zero modes
(among which many terms cancel).
The zero modes will provide reliable results
to leading order in $\lambda$, but will also contribute
to subleading terms. However, subleading terms can and
will receive contributions from other modes.
Hence, we will only collect the leading order
result.
We also take into account that $\vec{q}_1$ and $\vec{q}_2$
are dummy variables that can be interchanged in individual terms.
Terms which do not depend on both $n_1$ and $n_2$ vanish 
in zeta function regularization.
Taking all this into account, the integrands become
\begin{equation}\nonumber
\begin{split}
  I_6^0(\vec{q}_1,\vec{q}_2) & = 
    -42 \Delta_S(\vec{q}_1) \Delta_S(\vec{q}_2), \\
  I_{44}^0(\vec{q}_1,\vec{q}_2) & = 0, \\ 
  I_{433}^0(\vec{q}_1,\vec{q}_2) & =
\frac{8}{3} 
\left[
2 \vec{q}_1^2 \vec{q}_1^2 
-2 \vec{q}_1^2 \vec{q}_2^2 
- (\vec{q}_1 + \vec{q}_2)^2 (\vec{q}_1 + \vec{q}_2)^2
\right]
\Delta_A(\vec{q}_1 + \vec{q}_2)^2 
\Delta(\vec{q}_{f1}) \Delta(\vec{q}_{f2}) 
+ \\
& +
\frac{8}{3} 
\left[
4 \vec{q}_1^2 (\vec{q}_1 + \vec{q}_2)^2
-2 \vec{q}_1^2 \vec{q}_1^2 
- (\vec{q}_1 + \vec{q}_2)^2 (\vec{q}_1 + \vec{q}_2)^2
\right]
\Delta_A(\vec{q}_1 + \vec{q}_2)^2 
\Delta_S(\vec{q}_1) \Delta_S(\vec{q}_2),
\\
  I_{3333}^0(\vec{q}_1,\vec{q}_2) & = 0,
\end{split}
\end{equation}
still written in the notation of (\ref{eq_shortprop}), (\ref{eq_not1}) and (\ref{eq_not2}).
We still need to do the integrals over $\vec{q}_1$ and $\vec{q}_2$.
The theory can be 
regularized in the
ultraviolet (finite temperature only changes the theory at
scales $\gtrsim \frac{1}{T}$). Thus, the leading order result can be collected from
the IR divergent end of the integrals. The result is
\begin{equation}
\begin{split}
-\lim_{\vec{p}\rightarrow 0} & M^2(\vec{p},0)
=
  2\pi k T^2 \lambda^2 
    \int \frac{d^2 q_1}{(2\pi)^2}   
    \int \frac{d^2 q_2}{(2\pi)^2}  
\left[
  I_{6}^0(\vec{q_1},\vec{q}_2)
+ I_{433}^0(\vec{q_1},\vec{q}_2)
\right]
+ \text{higher modes} = \\
& =
  2\pi k T^2 \lambda^2
\left[
  -42 \frac{1}{(2\pi)^2} \log(\mu)^2
+
  \frac{16}{3} \frac{1}{(2\pi)^2} \log(\mu)^2
-
  \frac{8}{3} \frac{1}{(2\pi)^2} \log(\mu)^2
\right]
+ 
\mathcal{O}(\lambda^2 \log(\lambda)) = \\
& =
  -2\pi k T^2 \lambda^2
\left[
  \frac{118}{3(2\pi)^2} \log(\mu)^2
\right]
+ 
\mathcal{O}(\lambda^2 \log(\lambda)).
\end{split}
\end{equation}
Comparing to (\ref{eq_defM}), we identify
\begin{equation}\label{eq_scalar_selfenergy}
\begin{split}
  \mu^2(\lambda) 
& = \frac{118}{3 (2\pi)^2} 
  \lambda^2 \log(\mu)^2 
  + 
  \mathcal{O}(\lambda^2 \log(\lambda)).
\end{split}
\end{equation}
Notice that the right-hand side itself depends on $\mu^2(\lambda)$.
This is due to the fact that 
consistency (IR finiteness) requires that we use
renormalized propagators (\ref{eq_rens})
when we compute the values of the self-energy diagrams
\cite{D'Hoker:1981us}.
%
Summing up the one-particle-irreducible contributions in the
conventional geometric series
\begin{eqnarray}\nonumber
    \raisebox{-9mm}{
    \begin{fmffile}{diagselfenergys1}
    \begin{fmfgraph}(60,60)
       \fmfleft{i1}
       \fmfright{o1}
       \fmfblob{.30w}{v1}
       \fmf{plain}{o1,v1,i1}
       \fmfdot{i1,o1}
    \end{fmfgraph}
    \end{fmffile}}
=
    \raisebox{-9mm}{
    \begin{fmffile}{diagselfenergys2}
    \begin{fmfgraph}(60,60)
       \fmfleft{i1}
       \fmfright{o1}
       \fmf{plain}{o1,i1}
       \fmfdot{i1,o1}
    \end{fmfgraph}
    \end{fmffile}}
+
    \raisebox{-9mm}{
    \begin{fmffile}{diagselfenergys3}
    \begin{fmfgraph}(60,60)
       \fmfleft{i1}
       \fmfright{o1}
       \fmfv{decor.shape=circle,decor.filled=empty}{v1}
       \fmf{plain}{o1,v1,i1}
       \fmfdot{i1,o1}
    \end{fmfgraph}
    \end{fmffile}}
+
    \raisebox{-9mm}{
    \begin{fmffile}{diagselfenergys4}
    \begin{fmfgraph}(60,60)
       \fmfleft{i1}
       \fmfright{o1}
       \fmfv{decor.shape=circle,decor.filled=empty}{v1}
       \fmfv{decor.shape=circle,decor.filled=empty}{v2}
       \fmf{plain}{o1,v1,v2,i1}
       \fmfdot{i1,o1}
    \end{fmfgraph}
    \end{fmffile}}
+
  \cdots, \\ \nonumber
\end{eqnarray}
we find the renormalized scalar two-point function (external lines are included)
\begin{equation}\label{eq_rens}
 \raisebox{-12mm}{
    \begin{fmffile}{diagselfenergyYY}
    \begin{fmfgraph}(75,75)
       \fmfleft{i1}
       \fmfright{o1}
       \fmfblob{.30w}{v1}
       \fmf{plain}{o1,v1,i1}
       \fmfdot{i1,o1}
    \end{fmfgraph}
    \end{fmffile}}
=
  \frac{2\pi}{k T} \, \frac{1}{\vec{p}^2+\omega_n^2+m_Y^2},
\end{equation}
where the scalar thermal mass is
\begin{equation}\label{eq_scalar_thermal_mass}
  m_Y^2(\lambda) = (2\pi T)^2\mu^2(\lambda),
\end{equation}
and $\mu^2(\lambda)$ is given by equation (\ref{eq_scalar_selfenergy}).


\section{Gluons}\label{app_gluons}

In this section, we compute the static self-energy of gluons and ghosts in Lorentz gauge\footnote{
Note that Coulomb gauge is our main gauge choice throughout all other parts of this note. This section is primarily meant to illustrate the problems with Lorentz gauge, in particular the non-existence of an IR regulating mass in a gauge which leads to propagating gluons and ghosts.
}. We will find that to two-loop order, no thermal mass is generated.

We first consider one-loop diagrams.
Using (\ref{eq_interactions}) and (\ref{eq_propagators}),
we find that the following one-particle-irreducible diagrams
contribute to the gluon self-energy in the static limit.
\begin{equation}\label{eq_AAcon1}
\raisebox{-17mm}{
\begin{fmffile}{diagAAscalar4}
    \begin{fmfgraph}(100,100)
       \fmfleft{i1}
       \fmfright{o1}
       \fmf{wiggly}{o1,v1,i1}
       \fmf{plain,right=0.5}{v1,v1}
       \fmfdot{v1}
    \end{fmfgraph}
\end{fmffile}}
=-8kT
 \lambda \delta_{ij}
\int \frac{d^2q}{(2\pi)^2} T \sum_{n \in \mathbb{Z}} 
    \Delta_S(q),
\end{equation}

\vspace{-20mm}

\begin{equation}\label{eq_AAcon2}
\raisebox{-17mm}{
\begin{fmffile}{diagAAscalar}
    \begin{fmfgraph}(100,100)
       \fmfleft{i1}
       \fmfright{o1}
       \fmf{wiggly}{o1,v1}
       \fmf{wiggly}{v2,i1}
       \fmf{plain,right=1.0,tension=0.75}{v1,v2}
       \fmf{plain,right=1.0,tension=0.75}{v2,v1}
    \end{fmfgraph}
\end{fmffile}}
=
16 kT
 \lambda
\int \frac{d^2q}{(2\pi)^2} T \sum_{n \in \mathbb{Z}} 
    q_i q_j \Delta_S(q)^2,
\end{equation}

\vspace{-20mm}

\begin{equation}\label{eq_AAcon3}
\raisebox{-17mm}{
\begin{fmffile}{diagAAfermion}
    \begin{fmfgraph}(100,100)
       \fmfleft{i1}
       \fmfright{o1}
       \fmf{wiggly}{o1,v1}
       \fmf{wiggly}{v2,i1}
       \fmf{dashes,right=1.0,tension=0.75}{v1,v2}
       \fmf{dashes,right=1.0,tension=0.75}{v2,v1}
    \end{fmfgraph}
\end{fmffile}}
=
8kT
 \lambda 
\int \frac{d^2q}{(2\pi)^2} T \sum_{n \in \mathbb{Z}+\frac{1}{2}} 
    \left( q^2\delta_{ij} - 2 q_i q_j \right) \Delta(q)^2,
\end{equation}

\begin{figure}[t]
 \centering 
    \begin{picture}(0,0)
{\large
    \put(110, 153){\makebox(0,0)[l]{$\text{t}$}}
    \put(43, 112){\makebox(0,0)[l]{$\text{c}$}}
    \put(78, 23){\makebox(0,0)[l]{$\text{b}$}}
}
    \end{picture}
  \includegraphics{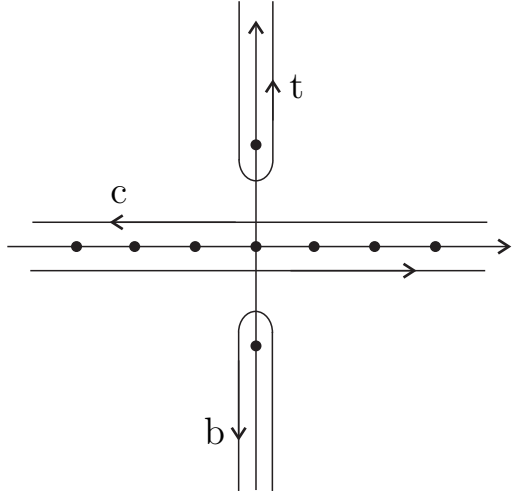}
  \caption{Integration contours and poles.}\label{fig_contour}
\end{figure}

\vspace{-20mm}

\begin{equation}\label{eq_AAcon4}
\raisebox{-17mm}{
\begin{fmffile}{diagAAgluon}
    \begin{fmfgraph}(100,100)
       \fmfleft{i1}
       \fmfright{o1}
       \fmf{wiggly}{o1,v1}
       \fmf{wiggly}{v2,i1}
       \fmf{wiggly,right=0.75,tension=0.6}{v1,v2}
       \fmf{wiggly,right=0.75,tension=0.6}{v2,v1}
    \end{fmfgraph}
\end{fmffile}}
+
\raisebox{-17mm}{
\begin{fmffile}{diagAAghost}
    \begin{fmfgraph}(100,100)
       \fmfleft{i1}
       \fmfright{o1}
       \fmf{wiggly}{o1,v1}
       \fmf{wiggly}{v2,i1}
       \fmf{dots,right=0.9,tension=0.75}{v1,v2}
       \fmf{dots,right=0.9,tension=0.75}{v2,v1}
    \end{fmfgraph}
\end{fmffile}}
= 0,
\end{equation}
written using the definitions in (\ref{eq_shortprop}).
The sum of gluon and ghost
loops cancel exactly if we use uncorrected propagators.
In principle, the result (\ref{eq_AAcon4})
may receive corrections from renormalized 
propagators \cite{D'Hoker:1981us}. 
It turns out that diagrams contributing to renormalization of
$A$ and $\hat{A}$ are equal.
We can sum up
(\ref{eq_AAcon1}), (\ref{eq_AAcon2}), (\ref{eq_AAcon3})
and (\ref{eq_AAcon4}) either by doing the sum or the integral first. If we do the sum first, the summation formulas
\begin{equation}\label{eq_spsums}
\begin{split}
  \sum_{n \in \mathbb{Z}} \frac{1}{n^2 + \Delta^2}
  & =  \frac{\pi \coth(\pi \Delta)}{\Delta}, \\
  \sum_{n \in \mathbb{Z}+\frac{1}{2}} \frac{1}{n^2 + \Delta^2}
  & = \frac{\pi \tanh(\pi \Delta)}{\Delta}, \\
\end{split}
\end{equation}
and
\begin{equation}\label{eq_dpsums}
\begin{split}
  & \sum_{n \in \mathbb{Z}} \frac{n^p}{\left(n^2 + \Delta^2\right)^2}=
\delta_{p,0}
\frac{\pi \coth(\pi \Delta)}{2 \Delta^3} +
\delta_{p,2}
\frac{\pi \coth(\pi \Delta)}{2 \Delta}, \\
  & \sum_{n \in \mathbb{Z}+\frac{1}{2}} \frac{n^p}{\left(n^2 + \Delta^2\right)^2} = 
 \delta_{p,0} \frac{\pi \tanh(\pi \Delta)}{2 \Delta^3} +
 \delta_{p,2} \frac{\pi \tanh(\pi \Delta)}{2 \Delta},
\end{split}
\end{equation}
(for $p \in \{ 0,1,2 \}$) are useful.
The formulas (\ref{eq_spsums}) and (\ref{eq_dpsums})
can be derived using a method similar to the one in \cite{Klebanov:2002mp}, by
considering the contour integrals
\begin{equation}\label{eq_contour}
\begin{split}
  0 & = I_\text{t}+I_\text{b}+I_\text{c} 
      = \oint \frac{dz}{2\pi i} f(z) \pi \cot(\pi z), \\
\vspace{30mm}   
  0 & = I_\text{t}+I_\text{b}+I_\text{c} 
      = \oint \frac{dz}{2\pi i} f(z) (-\pi) \tan(\pi z).
\end{split}
\end{equation}
Poles and integration contours $t$, $b$ and $c$ are
shown in figure \ref{fig_contour} (for the case with a pole at the origin).
Note that the function $\pi \cot(\pi z)$
has simple poles at $z=n$, $n\in \mathbb{Z}$,
and that the function
$-\pi \tan(\pi z)$
has simple poles at 
$z=n$, $n\in \mathbb{Z}+\frac{1}{2}$.
The summation formulas (\ref{eq_spsums}) and 
(\ref{eq_dpsums}) then follow
by setting $f(z) \rightarrow g_1(z)$ and
$f(z) \rightarrow g_2(z)$, respectively, 
in (\ref{eq_contour}). The functions $g_1(z)$ and $g_2(z)$ are defined by
\begin{align}
  g_1(z) & = \frac{1}{z^2+\Delta^2}, \\
  g_2(z) & = \frac{z^p}{\left(z^2+\Delta^2\right)^2},
\end{align}
where $p \in \{0,1,2\}$.
The functions $g_1(z)$ and $g_2(z)$ have simple and double poles, respectively, 
at $z=\pm i \Delta$.


Either way, 
(\ref{eq_AAcon1}), (\ref{eq_AAcon2}), (\ref{eq_AAcon3})
and (\ref{eq_AAcon4}) sum up to\footnote{
I thank J. Maldacena for discussions on the expected form of the gluon self-energy.
}
\begin{equation}
\raisebox{-17mm}{
\begin{fmffile}{diagAAtotal}
    \begin{fmfgraph}(100,100)
       \fmfleft{i1}
       \fmfright{o1}
       \fmfv{decor.shape=circle,decor.filled=empty}{v1}
       \fmf{wiggly}{o1,v1,i1}
    \end{fmfgraph}
\end{fmffile}} 
=\lim_{\vec{p}\rightarrow 0} \Pi_{ij}(\vec{p},0)
=\delta_{i,3} \delta_{j,3} k T^2 \Pi(\lambda,\mu),
\end{equation}
where $\Pi(\lambda,\mu) = \frac{8 \lambda}{2\pi} \log(\mu)$
(to lowest order in $\lambda$).
Hence, the gluon two-point function becomes
(upper sign corresponds to $A_i$, lower sign to $\hat{A}_i$)
\begin{equation}\label{eq_oneloop}
\begin{split}
\hspace{-100mm}
&   \raisebox{-10mm}{
    \begin{fmffile}{diagselfenergy1}
    \begin{fmfgraph}(60,60)
       \fmfleft{i1}
       \fmfright{o1}
       \fmfblob{.30w}{v1}
       \fmf{wiggly}{o1,v1,i1}
       \fmfdot{i1,o1}
    \end{fmfgraph}
    \end{fmffile}}
=  \left[ \frac{kT}{2\pi} \left( \pm \levi_{ijk} p_k + \frac{1}{2\xi} p_i p_j \right) + \delta_{i3} \delta_{j3} kT^2 \Pi\right]^{-1} \rightarrow \\
& \rightarrow (D_1)_{ij}(p)=
(D_0)_{ij}(p)+\frac{(2\pi)^2}{(kT)^2} kT^2\Pi
\frac{1}{(p^2)^2}
\begin{pmatrix}
  p_2^2 & -p_1 p_2 & 0  \\
  - p_1 p_2 & p_1^2 & 0  \\
  0 & 0 & 0 \\
\end{pmatrix}  + \mathcal{O}(\lambda^2),
\end{split}
\end{equation}
where the tree-level propagator is
\begin{equation}\label{eq_tree}
  \left(\frac{kT}{2\pi}\right)^{-1} \left[ \pm \levi_{ijk} p_k + \frac{1}{2\xi} p_i p_j \right]^{-1} \rightarrow  (D_0)_{ij}(p) = \pm \frac{2\pi}{kT}
\frac{1}{p^2}
\begin{pmatrix}
  0 & -p_3 & p_2  \\
  p_3 & 0 & -p_1  \\
  -p_2 & p_1 & 0 \\
\end{pmatrix},
\end{equation}
and the arrow denotes the operation of inverting the operator and then using the Landau gauge choice\footnote{
The corresponding result in Coulomb gauge is obtained by setting $p_3 \rightarrow 0$.
}
$\xi = 0$.


Comparing (\ref{eq_oneloop}) to
(\ref{eq_tree}), we note that no mass is generated and the correction is not sufficient to suppress the IR divergences.
Hence, we need to proceed to higher orders to find an IR regulator for the gluon, if it exists.
At two-loop order, only two diagrams can contribute for combinatorical reasons.
    \begin{eqnarray}\nonumber
&
    \raisebox{-12mm}{
    \begin{fmffile}{diagAA2loop1}
    \begin{fmfgraph}(80,80)
           \fmfleft{i}
           \fmfright{o}
           \fmf{phantom}{i,v1,v4,v2,v3,o}
           \fmffreeze
           \fmf{wiggly}{i,v1}
           \fmf{wiggly,left=0.4}{v1,v2}
           \fmf{plain,right=0.4}{v1,v2}
           \fmf{plain}{v2,v3}
           \fmf{plain,right=0.8}{v1,v3}
           \fmf{wiggly}{v3,o}
    \end{fmfgraph}
    \end{fmffile}}
+
    \raisebox{-12mm}{
    \begin{fmffile}{diagAA2loop2}
    \begin{fmfgraph}(80,80)
           \fmfleft{i}
           \fmfright{o}
           \fmf{phantom}{i,v1,v4,v2,v3,o}
           \fmffreeze
           \fmf{wiggly}{i,v1}
           \fmf{plain,left=0.4}{v1,v2}
           \fmf{wiggly,right=0.4}{v1,v2}
           \fmf{plain}{v2,v3}
           \fmf{plain,right=0.8}{v1,v3}
           \fmf{wiggly}{v3,o}
    \end{fmfgraph}
    \end{fmffile}} 
=
  \int \frac{d^2q_1}{(2\pi)^2} T \sum_{n_1 \in \mathbb{Z}}
  \int \frac{d^2q_2}{(2\pi)^2} T \sum_{n_2 \in \mathbb{Z}}
  R_{ij}=0,
\\
  & R_{ij} = 2 \pi kT \lambda^2
  \frac{64}{3} \left[ (q_2)_i \levi_{jkl} + (q_2)_j \levi_{ikl} \right]
  (q_1)_k (q_2)_l 
  \Delta_A(q_1) \Delta_S(q_2)^2 \Delta_S(q_1+q_2),
\label{eq_gluon2loop}
\end{eqnarray}
where $q_1 = (\vec{q_1},2 \pi T n_1)$ and $q_2 = (\vec{q_2},2 \pi T n_2)$.
The last equality sign comes from doing the loop integrals. Hence, the static self-energy vanishes at two-loop order.


Ghost self-energy diagrams always contain
the vertex structure $V_3^{\text{gh}}$
(as defined in (\ref{eq_interactions})).
Using this fact and the static limit,
it can be verified that to two-loop order,
no thermal mass is generated
for the ghosts, 
\begin{equation}
  m_c^2=0.
\end{equation}





\bibliographystyle{nb}
\bibliography{entropy}

\begin{thebibliography}{10}
\ifx\href\asklfhas\newcommand{\href}[2]{#2}\fi
\ifx\arxivref\asklfhas\newcommand{\arxivref}[1]{\href{http://arxiv.org/abs/#1}%
{#1}}\fi
\ifx\doiref\asklfhas\newcommand{\doiref}[2]{\href{http://dx.doi.org/#1}{#2}}\fi
\raggedright
\small
\parskip 0pt

\bibitem{Maldacena:1997re}
J.~M.~Maldacena,
\textit{``{The large N limit of superconformal field theories and
  supergravity}''},
\textsf{Adv.~Theor.~Math.~Phys.~2,~231~(1998)},
\texttt{\arxivref{hep-th/9711200}}.
%
\bibitem{Gubser:1998bc}
S.~S.~Gubser, I.~R.~Klebanov and A.~M.~Polyakov,
\textit{``{Gauge theory correlators from non-critical string theory}''},
\textsf{\doiref{10.1016/S0370-2693(98)00377-3}{Phys.~Lett.~B428,~105~(1998)}},
\texttt{\arxivref{hep-th/9802109}}.
%
\bibitem{Witten:1998qj}
E.~Witten,
\textit{``{Anti-de Sitter space and holography}''},
\textsf{Adv.~Theor.~Math.~Phys.~2,~253~(1998)},
\texttt{\arxivref{hep-th/9802150}}.
%
\bibitem{Bagger:2006sk}
J.~Bagger and N.~Lambert,
\textit{``{Modeling multiple M2's}''},
\textsf{\doiref{10.1103/PhysRevD.75.045020}{Phys.~Rev.~D75,~045020~(2007)}},
\texttt{\arxivref{hep-th/0611108}}.
%
\bibitem{Bagger:2007jr}
J.~Bagger and N.~Lambert,
\textit{``{Gauge Symmetry and Supersymmetry of Multiple M2-Branes}''},
\textsf{\doiref{10.1103/PhysRevD.77.065008}{Phys.~Rev.~D77,~065008~(2008)}},
\texttt{\arxivref{0711.0955}}.
%
\bibitem{Bagger:2007vi}
J.~Bagger and N.~Lambert,
\textit{``{Comments On Multiple M2-branes}''},
\textsf{\doiref{10.1088/1126-6708/2008/02/105}{JHEP~0802,~105~(2008)}},
\texttt{\arxivref{0712.3738}}.
%
\bibitem{Gustavsson:2007vu}
A.~Gustavsson,
\textit{``{Algebraic structures on parallel M2-branes}''},
\textsf{\doiref{10.1016/j.nuclphysb.2008.11.014}{Nucl.~Phys.~B811,~66~(2009)}},
\texttt{\arxivref{0709.1260}}.
%
\bibitem{Schwarz:2004yj}
J.~H.~Schwarz,
\textit{``{Superconformal Chern-Simons theories}''},
\textsf{\doiref{10.1088/1126-6708/2004/11/078}{JHEP~0411,~078~(2004)}},
\texttt{\arxivref{hep-th/0411077}}.
%
\bibitem{VanRaamsdonk:2008ft}
M.~Van~Raamsdonk,
\textit{``{Comments on the Bagger-Lambert theory and multiple M2- branes}''},
\textsf{\doiref{10.1088/1126-6708/2008/05/105}{JHEP~0805,~105~(2008)}},
\texttt{\arxivref{0803.3803}}.
%
\bibitem{Bandres:2008vf}
M.~A.~Bandres, A.~E.~Lipstein and J.~H.~Schwarz,
\textit{``{N = 8 Superconformal Chern--Simons Theories}''},
\textsf{\doiref{10.1088/1126-6708/2008/05/025}{JHEP~0805,~025~(2008)}},
\texttt{\arxivref{0803.3242}}.
%
\bibitem{Lambert:2008et}
N.~Lambert and D.~Tong,
\textit{``{Membranes on an Orbifold}''},
\textsf{\doiref{10.1103/PhysRevLett.101.041602}{Phys.~Rev.~Lett.~101,~041602~(%
2008)}},
\texttt{\arxivref{0804.1114}}.
%
\bibitem{Distler:2008mk}
J.~Distler, S.~Mukhi, C.~Papageorgakis and M.~Van~Raamsdonk,
\textit{``{M2-branes on M-folds}''},
\textsf{\doiref{10.1088/1126-6708/2008/05/038}{JHEP~0805,~038~(2008)}},
\texttt{\arxivref{0804.1256}}.
%
\bibitem{Gran:2008vi}
U.~Gran, B.~E.~W.~Nilsson and C.~Petersson,
\textit{``{On relating multiple M2 and D2-branes}''},
\textsf{\doiref{10.1088/1126-6708/2008/10/067}{JHEP~0810,~067~(2008)}},
\texttt{\arxivref{0804.1784}}.
%
\bibitem{Gomis:2008uv}
J.~Gomis, G.~Milanesi and J.~G.~Russo,
\textit{``{Bagger-Lambert Theory for General Lie Algebras}''},
\textsf{\doiref{10.1088/1126-6708/2008/06/075}{JHEP~0806,~075~(2008)}},
\texttt{\arxivref{0805.1012}}.
%
\bibitem{Benvenuti:2008bt}
S.~Benvenuti, D.~Rodriguez-Gomez, E.~Tonni and H.~Verlinde,
\textit{``{N=8 superconformal gauge theories and M2 branes}''},
\textsf{\doiref{10.1088/1126-6708/2009/01/078}{JHEP~0901,~078~(2009)}},
\texttt{\arxivref{0805.1087}}.
%
\bibitem{Ho:2008ei}
P.-M.~Ho, Y.~Imamura and Y.~Matsuo,
\textit{``{M2 to D2 revisited}''},
\textsf{\doiref{10.1088/1126-6708/2008/07/003}{JHEP~0807,~003~(2008)}},
\texttt{\arxivref{0805.1202}}.
%
\bibitem{Bandres:2008kj}
M.~A.~Bandres, A.~E.~Lipstein and J.~H.~Schwarz,
\textit{``{Ghost-Free Superconformal Action for Multiple M2-Branes}''},
\textsf{\doiref{10.1088/1126-6708/2008/07/117}{JHEP~0807,~117~(2008)}},
\texttt{\arxivref{0806.0054}}.
%
\bibitem{Gomis:2008be}
J.~Gomis, D.~Rodriguez-Gomez, M.~Van~Raamsdonk and H.~Verlinde,
\textit{``{Supersymmetric Yang-Mills Theory From Lorentzian Three-
  Algebras}''},
\textsf{\doiref{10.1088/1126-6708/2008/08/094}{JHEP~0808,~094~(2008)}},
\texttt{\arxivref{0806.0738}}.
%
\bibitem{Aharony:2008ug}
O.~Aharony, O.~Bergman, D.~L.~Jafferis and J.~Maldacena,
\textit{``{N=6 superconformal Chern-Simons-matter theories, M2-branes and their
  gravity duals}''},
\textsf{\doiref{10.1088/1126-6708/2008/10/091}{JHEP~0810,~091~(2008)}},
\texttt{\arxivref{0806.1218}}.
%
\bibitem{Gaiotto:2007qi}
D.~Gaiotto and X.~Yin,
\textit{``{Notes on superconformal Chern-Simons-matter theories}''},
\textsf{\doiref{10.1088/1126-6708/2007/08/056}{JHEP~0708,~056~(2007)}},
\texttt{\arxivref{0704.3740}}.
%
\bibitem{Gaiotto:2008sd}
D.~Gaiotto and E.~Witten,
\textit{``{Janus Configurations, Chern-Simons Couplings, And The Theta-Angle in
  N=4 Super Yang-Mills Theory}''},
\texttt{\arxivref{0804.2907}}.
%
\bibitem{Hosomichi:2008jd}
K.~Hosomichi, K.-M.~Lee, S.~Lee, S.~Lee and J.~Park,
\textit{``{N=4 Superconformal Chern-Simons Theories with Hyper and Twisted
  Hyper Multiplets}''},
\textsf{\doiref{10.1088/1126-6708/2008/07/091}{JHEP~0807,~091~(2008)}},
\texttt{\arxivref{0805.3662}}.
%
\bibitem{Gauntlett:2008uf}
J.~P.~Gauntlett and J.~B.~Gutowski,
\textit{``{Constraining Maximally Supersymmetric Membrane Actions}''},
\textsf{JHEP~0806,~053~(2008)},
\texttt{\arxivref{0804.3078}}.
%
\bibitem{Papadopoulos:2008sk}
G.~Papadopoulos,
\textit{``{M2-branes, 3-Lie Algebras and Plucker relations}''},
\textsf{\doiref{10.1088/1126-6708/2008/05/054}{JHEP~0805,~054~(2008)}},
\texttt{\arxivref{0804.2662}}.
%
\bibitem{Benna:2008zy}
M.~Benna, I.~Klebanov, T.~Klose and M.~Smedb{\"a}ck,
\textit{``{Superconformal Chern-Simons Theories and AdS$_4$/CFT$_3$
  Correspondence}''},
\texttt{\arxivref{0806.1519}}.
%
\bibitem{Bandres:2008ry}
M.~A.~Bandres, A.~E.~Lipstein and J.~H.~Schwarz,
\textit{``{Studies of the ABJM Theory in a Formulation with Manifest SU(4)
  R-Symmetry}''},
\textsf{\doiref{10.1088/1126-6708/2008/09/027}{JHEP~0809,~027~(2008)}},
\texttt{\arxivref{0807.0880}}.
%
\bibitem{Berenstein:2008dc}
D.~Berenstein and D.~Trancanelli,
\textit{``{Three-dimensional N=6 SCFT's and their membrane dynamics}''},
\textsf{\doiref{10.1103/PhysRevD.78.106009}{Phys.~Rev.~D78,~106009~(2008)}},
\texttt{\arxivref{0808.2503}}.
%
\bibitem{Klebanov:2008vq}
I.~Klebanov, T.~Klose and A.~Murugan,
\textit{``{$AdS_4/CFT_3$ -- Squashed, Stretched and Warped}''},
\textsf{\doiref{10.1088/1126-6708/2009/03/140}{JHEP~0903,~140~(2009)}},
\texttt{\arxivref{0809.3773}}.
%
\bibitem{Park:2008bk}
C.-S.~Park,
\textit{``{Comments on Baryon-like Operators in N=6 Chern-Simons- matter theory
  of ABJM}''},
\texttt{\arxivref{0810.1075}}.
%
\bibitem{Imamura:2009ur}
Y.~Imamura,
\textit{``{Monopole operators in N=4 Chern-Simons theories and wrapped
  M2-branes}''},
\textsf{\doiref{10.1143/PTP.121.1173}{Prog.~Theor.~Phys.~121,~1173~(2009)}},
\texttt{\arxivref{0902.4173}}.
%
\bibitem{Gaiotto:2009tk}
D.~Gaiotto and D.~L.~Jafferis,
\textit{``{Notes on adding D6 branes wrapping RP(3) in AdS(4) x CP(3)}''},
\texttt{\arxivref{0903.2175}}.
%
\bibitem{SheikhJabbari:2009kr}
M.~M.~Sheikh-Jabbari and J.~Simon,
\textit{``{On Half-BPS States of the ABJM Theory}''},
\textsf{\doiref{10.1088/1126-6708/2009/08/073}{JHEP~0908,~073~(2009)}},
\texttt{\arxivref{0904.4605}}.
%
\bibitem{Benna:2009xd}
M.~K.~Benna, I.~R.~Klebanov and T.~Klose,
\textit{``{Charges of Monopole Operators in Chern-Simons Yang-Mills Theory}''},
\texttt{\arxivref{0906.3008}}.
%
\bibitem{Gustavsson:2009pm}
A.~Gustavsson and S.-J.~Rey,
\textit{``{Enhanced N=8 Supersymmetry of ABJM Theory on R(8) and R(8)/Z(2)}''},
\texttt{\arxivref{0906.3568}}.
%
\bibitem{Berenstein:2009sa}
D.~Berenstein and J.~Park,
\textit{``{The BPS spectrum of monopole operators in ABJM: towards a field
  theory description of the giant torus}''},
\texttt{\arxivref{0906.3817}}.
%
\bibitem{Kwon:2009ar}
O.-K.~Kwon, P.~Oh and J.~Sohn,
\textit{``{Notes on Supersymmetry Enhancement of ABJM Theory}''},
\textsf{\doiref{10.1088/1126-6708/2009/08/093}{JHEP~0908,~093~(2009)}},
\texttt{\arxivref{0906.4333}}.
%
\bibitem{Kim:2009ia}
S.~Kim and K.~Madhu,
\textit{``{Aspects of monopole operators in N=6 Chern-Simons theory}''},
\textsf{\doiref{10.1088/1126-6708/2009/12/018}{JHEP~0912,~018~(2009)}},
\texttt{\arxivref{0906.4751}}.
%
\bibitem{Imamura:2009hc}
Y.~Imamura and S.~Yokoyama,
\textit{``{A Monopole Index for N=4 Chern-Simons Theories}''},
\textsf{\doiref{10.1016/j.nuclphysb.2009.10.025}{Nucl.~Phys.~B827,~183~(2010)}%
},
\texttt{\arxivref{0908.0988}}.
%
\bibitem{Kwon:2010ev}
O.-K.~Kwon, P.~Oh, C.~Sochichiu and J.~Sohn,
\textit{``{Enhanced Supersymmetry of Nonrelativistic ABJM Theory}''},
\texttt{\arxivref{1001.1598}}.
%
\bibitem{Bhattacharya:2008bja}
J.~Bhattacharya and S.~Minwalla,
\textit{``{Superconformal Indices for ${\cal N}=6$ Chern Simons Theories}''},
\textsf{\doiref{10.1088/1126-6708/2009/01/014}{JHEP~0901,~014~(2009)}},
\texttt{\arxivref{0806.3251}}.
%
\bibitem{Kim:2009wb}
S.~Kim,
\textit{``{The complete superconformal index for N=6 Chern-Simons theory}''},
\textsf{\doiref{10.1016/j.nuclphysb.2009.06.025}{Nucl.~Phys.~B821,~241~(2009)}%
},
\texttt{\arxivref{0903.4172}}.
%
\bibitem{Lambert:2010ji}
N.~Lambert and C.~Papageorgakis,
\textit{``{Relating U(N)xU(N) to SU(N)xSU(N) Chern-Simons Membrane
  theories}''},
\texttt{\arxivref{1001.4779}}.
%
\bibitem{Nishioka:2008gz}
T.~Nishioka and T.~Takayanagi,
\textit{``{On Type IIA Penrose Limit and N=6 Chern-Simons Theories}''},
\textsf{\doiref{10.1088/1126-6708/2008/08/001}{JHEP~0808,~001~(2008)}},
\texttt{\arxivref{0806.3391}}.
%
\bibitem{Minahan:2008hf}
J.~A.~Minahan and K.~Zarembo,
\textit{``{The Bethe ansatz for superconformal Chern-Simons}''},
\textsf{\doiref{10.1088/1126-6708/2008/09/040}{JHEP~0809,~040~(2008)}},
\texttt{\arxivref{0806.3951}}.
%
\bibitem{Gaiotto:2008cg}
D.~Gaiotto, S.~Giombi and X.~Yin,
\textit{``{Spin Chains in N=6 Superconformal Chern-Simons-Matter Theory}''},
\texttt{\arxivref{0806.4589}}.
%
\bibitem{Grignani:2008is}
G.~Grignani, T.~Harmark and M.~Orselli,
\textit{``{The SU(2) x SU(2) sector in the string dual of N=6 superconformal
  Chern-Simons theory}''},
\textsf{\doiref{10.1016/j.nuclphysb.2008.10.019}{Nucl.~Phys.~B810,~115~(2009)}%
},
\texttt{\arxivref{0806.4959}}.
%
\bibitem{Gromov:2008bz}
N.~Gromov and P.~Vieira,
\textit{``{The AdS4/CFT3 algebraic curve}''},
\textsf{\doiref{10.1088/1126-6708/2009/02/040}{JHEP~0902,~040~(2009)}},
\texttt{\arxivref{0807.0437}}.
%
\bibitem{Gromov:2008qe}
N.~Gromov and P.~Vieira,
\textit{``{The all loop AdS4/CFT3 Bethe ansatz}''},
\textsf{\doiref{10.1088/1126-6708/2009/01/016}{JHEP~0901,~016~(2009)}},
\texttt{\arxivref{0807.0777}}.
%
\bibitem{Ahn:2008aa}
C.~Ahn and R.~I.~Nepomechie,
\textit{``{N=6 super Chern-Simons theory S-matrix and all-loop Bethe ansatz
  equations}''},
\textsf{\doiref{10.1088/1126-6708/2008/09/010}{JHEP~0809,~010~(2008)}},
\texttt{\arxivref{0807.1924}}.
%
\bibitem{Bak:2008cp}
D.~Bak and S.-J.~Rey,
\textit{``{Integrable Spin Chain in Superconformal Chern-Simons Theory}''},
\textsf{\doiref{10.1088/1126-6708/2008/10/053}{JHEP~0810,~053~(2008)}},
\texttt{\arxivref{0807.2063}}.
%
\bibitem{McLoughlin:2008he}
T.~McLoughlin, R.~Roiban and A.~A.~Tseytlin,
\textit{``{Quantum spinning strings in $AdS_4 \times CP^3$: testing the Bethe
  Ansatz proposal}''},
\textsf{\doiref{10.1088/1126-6708/2008/11/069}{JHEP~0811,~069~(2008)}},
\texttt{\arxivref{0809.4038}}.
%
\bibitem{Kristjansen:2008ib}
C.~Kristjansen, M.~Orselli and K.~Zoubos,
\textit{``{Non-planar ABJM Theory and Integrability}''},
\textsf{\doiref{10.1088/1126-6708/2009/03/037}{JHEP~0903,~037~(2009)}},
\texttt{\arxivref{0811.2150}}.
%
\bibitem{Minahan:2009te}
J.~A.~Minahan, W.~Schulgin and K.~Zarembo,
\textit{``{Two loop integrability for Chern-Simons theories with N=6
  supersymmetry}''},
\texttt{\arxivref{0901.1142}}.
%
\bibitem{Berenstein:2009qd}
D.~Berenstein and D.~Trancanelli,
\textit{``{S-duality and the giant magnon dispersion relation}''},
\texttt{\arxivref{0904.0444}}.
%
\bibitem{Bak:2009mq}
D.~Bak, H.~Min and S.-J.~Rey,
\textit{``{Generalized Dynamical Spin Chain and 4-Loop Integrability in N=6
  Superconformal Chern-Simons Theory}''},
\textsf{\doiref{10.1016/j.nuclphysb.2009.10.011}{Nucl.~Phys.~B827,~381~(2010)}%
},
\texttt{\arxivref{0904.4677}}.
%
\bibitem{Bak:2009tq}
D.~Bak, H.~Min and S.-J.~Rey,
\textit{``{Integrability of N=6 Chern-Simons Theory at Six Loops and
  Beyond}''},
\texttt{\arxivref{0911.0689}}.
%
\bibitem{Minahan:2009aq}
J.~A.~Minahan, O.~O.~Sax and C.~Sieg,
\textit{``{Magnon dispersion to four loops in the ABJM and ABJ models}''},
\texttt{\arxivref{0908.2463}}.
%
\bibitem{Minahan:2009wg}
J.~A.~Minahan, O.~O.~Sax and C.~Sieg,
\textit{``{Anomalous dimensions at four loops in N=6 superconformal
  Chern-Simons theories}''},
\texttt{\arxivref{0912.3460}}.
%
\bibitem{KeskiVakkuri:2008eb}
E.~Keski-Vakkuri and P.~Kraus,
\textit{``{Quantum Hall Effect in AdS/CFT}''},
\textsf{\doiref{10.1088/1126-6708/2008/09/130}{JHEP~0809,~130~(2008)}},
\texttt{\arxivref{0805.4643}}.
%
\bibitem{Davis:2008nv}
J.~L.~Davis, P.~Kraus and A.~Shah,
\textit{``{Gravity Dual of a Quantum Hall Plateau Transition}''},
\textsf{\doiref{10.1088/1126-6708/2008/11/020}{JHEP~0811,~020~(2008)}},
\texttt{\arxivref{0809.1876}}.
%
\bibitem{Fujita:2009kw}
M.~Fujita, W.~Li, S.~Ryu and T.~Takayanagi,
\textit{``{Fractional Quantum Hall Effect via Holography: Chern- Simons, Edge
  States, and Hierarchy}''},
\textsf{\doiref{10.1088/1126-6708/2009/06/066}{JHEP~0906,~066~(2009)}},
\texttt{\arxivref{0901.0924}}.
%
\bibitem{Hikida:2009tp}
Y.~Hikida, W.~Li and T.~Takayanagi,
\textit{``{ABJM with Flavors and FQHE}''},
\textsf{\doiref{10.1088/1126-6708/2009/07/065}{JHEP~0907,~065~(2009)}},
\texttt{\arxivref{0903.2194}}.
%
\bibitem{Alanen:2009cn}
J.~Alanen, E.~Keski-Vakkuri, P.~Kraus and V.~Suur-Uski,
\textit{``{AC Transport at Holographic Quantum Hall Transitions}''},
\texttt{\arxivref{0905.4538}}.
%
\bibitem{Gomis:2008vc}
J.~Gomis, D.~Rodriguez-Gomez, M.~Van~Raamsdonk and H.~Verlinde,
\textit{``{A Massive Study of M2-brane Proposals}''},
\textsf{\doiref{10.1088/1126-6708/2008/09/113}{JHEP~0809,~113~(2008)}},
\texttt{\arxivref{0807.1074}}.
%
\bibitem{Gauntlett:2009zw}
J.~P.~Gauntlett, S.~Kim, O.~Varela and D.~Waldram,
\textit{``{Consistent supersymmetric Kaluza--Klein truncations with massive
  modes}''},
\textsf{\doiref{10.1088/1126-6708/2009/04/102}{JHEP~0904,~102~(2009)}},
\texttt{\arxivref{0901.0676}}.
%
\bibitem{Denef:2009tp}
F.~Denef and S.~A.~Hartnoll,
\textit{``{Landscape of superconducting membranes}''},
\textsf{\doiref{10.1103/PhysRevD.79.126008}{Phys.~Rev.~D79,~126008~(2009)}},
\texttt{\arxivref{0901.1160}}.
%
\bibitem{Gauntlett:2009dn}
J.~P.~Gauntlett, J.~Sonner and T.~Wiseman,
\textit{``{Holographic superconductivity in M-Theory}''},
\textsf{\doiref{10.1103/PhysRevLett.103.151601}{Phys.~Rev.~Lett.~103,~151601~(%
2009)}},
\texttt{\arxivref{0907.3796}}.
%
\bibitem{Gauntlett:2009bh}
J.~Gauntlett, J.~Sonner and T.~Wiseman,
\textit{``{Quantum Criticality and Holographic Superconductors in M-
  theory}''},
\textsf{\doiref{10.1007/JHEP02(2010)060}{JHEP~1002,~060~(2010)}},
\texttt{\arxivref{0912.0512}}.
%
\bibitem{Bak:2010yd}
D.~Bak and S.~Yun,
\textit{``{Thermal Aspects of ABJM theory: Currents and Condensations}''},
\texttt{\arxivref{1001.4089}}.
%
\bibitem{Witten:1998zw}
E.~Witten,
\textit{``{Anti-de Sitter space, thermal phase transition, and confinement in
  gauge theories}''},
\textsf{Adv.~Theor.~Math.~Phys.~2,~505~(1998)},
\texttt{\arxivref{hep-th/9803131}}.
%
\bibitem{Fotopoulos:1998es}
A.~Fotopoulos and T.~R.~Taylor,
\textit{``{Comment on two-loop free energy in N = 4 supersymmetric Yang-Mills
  theory at finite temperature}''},
\textsf{\doiref{10.1103/PhysRevD.59.061701}{Phys.~Rev.~D59,~061701~(1999)}},
\texttt{\arxivref{hep-th/9811224}}.
%
\bibitem{Gubser:1996de}
S.~S.~Gubser, I.~R.~Klebanov and A.~W.~Peet,
\textit{``{Entropy and Temperature of Black 3-Branes}''},
\textsf{\doiref{10.1103/PhysRevD.54.3915}{Phys.~Rev.~D54,~3915~(1996)}},
\texttt{\arxivref{hep-th/9602135}}.
%
\bibitem{Klebanov:1996un}
I.~R.~Klebanov and A.~A.~Tseytlin,
\textit{``{Entropy of Near-Extremal Black p-branes}''},
\textsf{\doiref{10.1016/0550-3213(96)00295-7}{Nucl.~Phys.~B475,~164~(1996)}},
\texttt{\arxivref{hep-th/9604089}}.
%
\bibitem{Gubser:1998nz}
S.~S.~Gubser, I.~R.~Klebanov and A.~A.~Tseytlin,
\textit{``{Coupling constant dependence in the thermodynamics of N = 4
  supersymmetric Yang-Mills theory}''},
\textsf{\doiref{10.1016/S0550-3213(98)00514-8}{Nucl.~Phys.~B534,~202~(1998)}},
\texttt{\arxivref{hep-th/9805156}}.
%
\bibitem{VazquezMozo:1999ic}
M.~A.~Vazquez-Mozo,
\textit{``{A note on supersymmetric Yang-Mills thermodynamics}''},
\textsf{\doiref{10.1103/PhysRevD.60.106010}{Phys.~Rev.~D60,~106010~(1999)}},
\texttt{\arxivref{hep-th/9905030}}.
%
\bibitem{Kim:1999sg}
C.-j.~Kim and S.-J.~Rey,
\textit{``{Thermodynamics of large-N super Yang-Mills theory and AdS/CFT
  correspondence}''},
\textsf{\doiref{10.1016/S0550-3213(99)00532-5}{Nucl.~Phys.~B564,~430~(2000)}},
\texttt{\arxivref{hep-th/9905205}}.
%
\bibitem{Nieto:1999kc}
A.~Nieto and M.~H.~G.~Tytgat,
\textit{``{Effective field theory approach to N = 4 supersymmetric Yang-Mills
  at finite temperature}''},
\texttt{\arxivref{hep-th/9906147}}.
%
\bibitem{Garousi:2008ik}
M.~R.~Garousi, A.~Ghodsi and M.~Khosravi,
\textit{``{On thermodynamics of N=6 superconformal Chern-Simons theories at
  strong coupling}''},
\textsf{\doiref{10.1088/1126-6708/2008/08/067}{JHEP~0808,~067~(2008)}},
\texttt{\arxivref{0807.1478}}.
%
\bibitem{Klebanov:2009sg}
I.~R.~Klebanov and G.~Torri,
\textit{``{M2-branes and AdS/CFT}''},
\texttt{\arxivref{0909.1580}}.
%
\bibitem{Aharony:2008gk}
O.~Aharony, O.~Bergman and D.~L.~Jafferis,
\textit{``{Fractional M2-branes}''},
\textsf{\doiref{10.1088/1126-6708/2008/11/043}{JHEP~0811,~043~(2008)}},
\texttt{\arxivref{0807.4924}}.
%
\bibitem{Cederwall:2008xu}
M.~Cederwall,
\textit{``{Superfield actions for N=8 and N=6 conformal theories in three
  dimensions}''},
\textsf{\doiref{10.1088/1126-6708/2008/10/070}{JHEP~0810,~070~(2008)}},
\texttt{\arxivref{0809.0318}}.
%
\bibitem{Cederwall:2008vd}
M.~Cederwall,
\textit{``{N=8 superfield formulation of the Bagger-Lambert- Gustavsson
  model}''},
\textsf{\doiref{10.1088/1126-6708/2008/09/116}{JHEP~0809,~116~(2008)}},
\texttt{\arxivref{0808.3242}}.
%
\bibitem{Arnold:1994ps}
P.~Arnold and C.-X.~Zhai,
\textit{``{The three loop free energy for pure gauge QCD}''},
\textsf{\doiref{10.1103/PhysRevD.50.7603}{Phys.~Rev.~D50,~7603~(1994)}},
\texttt{\arxivref{hep-ph/9408276}}.
%
\bibitem{Arnold:1994eb}
P.~Arnold and C.-x.~Zhai,
\textit{``{The Three loop free energy for high temperature QED and QCD with
  fermions}''},
\textsf{\doiref{10.1103/PhysRevD.51.1906}{Phys.~Rev.~D51,~1906~(1995)}},
\texttt{\arxivref{hep-ph/9410360}}.
%
\bibitem{Kapusta:1979fh}
J.~I.~Kapusta,
\textit{``{Quantum Chromodynamics at High Temperature}''},
\textsf{\doiref{10.1016/0550-3213(79)90146-9}{Nucl.~Phys.~B148,~461~(1979)}}.
%
\bibitem{Gross:1980br}
D.~J.~Gross, R.~D.~Pisarski and L.~G.~Yaffe,
\textit{``{QCD and Instantons at Finite Temperature}''},
\textsf{\doiref{10.1103/RevModPhys.53.43}{Rev.~Mod.~Phys.~53,~43~(1981)}}.
%
\bibitem{D'Hoker:1981us}
E.~D'Hoker,
\textit{``{Perturbative results on QCD in three dimensions at finite
  temperature}''},
\textsf{\doiref{10.1016/0550-3213(82)90441-2}{Nucl.~Phys.~B201,~401~(1982)}}.
%
\bibitem{Kapusta:1989bd}
J.~I.~Kapusta and P.~V.~Landshoff,
\textit{``{Finite Temperature Field Theory}''},
\textsf{J.~Phys.~G15,~267~(1989)}.
%
\bibitem{Linde:1980ts}
A.~D.~Linde,
\textit{``{Infrared Problem in Thermodynamics of the Yang-Mills Gas}''},
\textsf{\doiref{10.1016/0370-2693(80)90769-8}{Phys.~Lett.~B96,~289~(1980)}}.
%
\bibitem{Li:1998kd}
M.~Li,
\textit{``{Evidence for large N phase transition in N = 4 super Yang- Mills
  theory at finite temperature}''},
\textsf{JHEP~9903,~004~(1999)},
\texttt{\arxivref{hep-th/9807196}}.
%
\bibitem{Gao:1998ww}
Y.-h.~Gao and M.~Li,
\textit{``{Large N strong/weak coupling phase transition and the correspondence
  principle}''},
\textsf{\doiref{10.1016/S0550-3213(99)00234-5}{Nucl.~Phys.~B551,~229~(1999)}},
\texttt{\arxivref{hep-th/9810053}}.
%
\bibitem{Burgess:1999vb}
C.~P.~Burgess, N.~R.~Constable and R.~C.~Myers,
\textit{``{The free energy of N = 4 superYang-Mills and the AdS/CFT
  correspondence}''},
\textsf{JHEP~9908,~017~(1999)},
\texttt{\arxivref{hep-th/9907188}}.
%
\bibitem{Aharony:1999ti}
O.~Aharony, S.~S.~Gubser, J.~M.~Maldacena, H.~Ooguri and Y.~Oz,
\textit{``{Large N field theories, string theory and gravity}''},
\textsf{\doiref{10.1016/S0370-1573(99)00083-6}{Phys.~Rept.~323,~183~(2000)}},
\texttt{\arxivref{hep-th/9905111}}.
%
\bibitem{Blaizot:2006tk}
J.~P.~Blaizot, E.~Iancu, U.~Kraemmer and A.~Rebhan,
\textit{``{Hard-thermal-loop entropy of supersymmetric Yang-Mills theories}''},
\textsf{JHEP~0706,~035~(2007)},
\texttt{\arxivref{hep-ph/0611393}}.
%
\bibitem{Sundborg:1999ue}
B.~Sundborg,
\textit{``{The Hagedorn Transition, Deconfinement and N=4 SYM Theory}''},
\textsf{\doiref{10.1016/S0550-3213(00)00044-4}{Nucl.~Phys.~B573,~349~(2000)}},
\texttt{\arxivref{hep-th/9908001}}.
%
\bibitem{Aharony:2003sx}
O.~Aharony, J.~Marsano, S.~Minwalla, K.~Papadodimas and M.~Van~Raamsdonk,
\textit{``{The Hagedorn / deconfinement phase transition in weakly coupled
  large N gauge theories}''},
\textsf{Adv.~Theor.~Math.~Phys.~8,~603~(2004)},
\texttt{\arxivref{hep-th/0310285}}.
%
\bibitem{Spradlin:2004pp}
M.~Spradlin and A.~Volovich,
\textit{``{A pendant for Polya: The one-loop partition function of N = 4 SYM on
  R x S(3)}''},
\textsf{\doiref{10.1016/j.nuclphysb.2005.01.007}{Nucl.~Phys.~B711,~199~(2005)}%
},
\texttt{\arxivref{hep-th/0408178}}.
%
\bibitem{Klebanov:2002mp}
I.~R.~Klebanov, M.~Spradlin and A.~Volovich,
\textit{``{New effects in gauge theory from pp-wave superstrings}''},
\textsf{\doiref{10.1016/S0370-2693(02)02841-1}{Phys.~Lett.~B548,~111~(2002)}},
\texttt{\arxivref{hep-th/0206221}}.
%
\end{thebibliography}

\end{document}